\documentclass[a4paper,12pt]{article}  %% important: a4paper first
\pdfoutput=1
\usepackage{mathptmx}
\usepackage{mysmallsec}
\usepackage[format=hang,justification=justified,labelfont=bf,labelsep=quad]{caption}
\usepackage{amssymb}
\usepackage{amsfonts}
\usepackage{amsmath}
\usepackage{graphicx}
\usepackage{hyperref} 
\def\ci#1{\hbox{$\bigcirc\llap{\hbox{\footnotesize\bf#1}\kern.3em}$}}
\def\proof{\noindent{\it Proof.\enspace}}
\def\endproof{\hfill\strut\nobreak\hfill\tombstone\par\medbreak}
\def\tombstone{\hbox{\lower.4pt\vbox{\hrule\hbox{\vrule
  \kern7.6pt\vrule height7.6pt}\hrule}\kern.5pt}}
\def\reals{{\mathbb R}}
\def\0{\hbox{\bf0}}
\def\1{\hbox{\bf1}}
\def\T{^{\top}}
\def\Pprime{P^{\kern.9pt\prime}}
\def\dualcyc{C_f^{\kern.7pt m}}
\def\CM#1#2{C_{#1}^{\kern.9pt #2}}
\def\C#1{\CM{#1}{m}}
\def\Pell{P^{\kern.5pt\ell}}
\def\Rell{R^{\kern.5pt\ell}}
\def\E{\mathbb{E}}
\def\reals{{\mathbb R}} 

\newdimen\einr
\def\abs#1{\par\hangafter=1\hangindent=\einr
  \noindent\hbox to\einr{\ignorespaces#1\hfill}\ignorespaces} 

\newtheorem{theorem}{Theorem}

\newtheorem{proposition}[theorem]{Proposition}
\def\proof{\noindent{\em Proof.\enspace}}

\def\endproof{\hfill\strut\nobreak\hfill\tombstone\par\medbreak}
\def\tombstone{\hbox{\lower.4pt\vbox{\hrule\hbox{\vrule
  \kern7.6pt\vrule height7.6pt}\hrule}\kern.5pt}}

\title{%
Unit Vector Games
}

\author{Rahul Savani%
\thanks{Department of Computer Science,
University of Liverpool,
Liverpool L69 3BX, United Kingdom.
Email: rahul.savani@liverpool.ac.uk
} \and
\and Bernhard von Stengel%
\thanks{Department of Mathematics,
London School of Economics, London WC2A 2AE, United Kingdom.
Email: stengel@nash.lse.ac.uk}
}

\date{February 15, 2016}

\begin{document}
\maketitle

\vskip0ex
{\small
Published in: International Journal of Economic Theory 12
(2016), 7--27. doi: 10.1111/ijet.12077

}
\vskip3ex

\begin{abstract}
\noindent
McLennan and Tourky (2010) showed that ``imitation games''
provide a new view of the computation of Nash
equilibria of bimatrix games with the Lemke--Howson
algorithm.
In an imitation game, the payoff matrix of one of the
players is the identity matrix.
We study the more general ``unit vector games'', which are
already known, where the payoff matrix of one player is
composed of unit vectors.
Our main application is a simplification of the construction
by Savani and von Stengel (2006) of bimatrix games where two
basic equilibrium-finding algorithms take exponentially many
steps: 
the Lemke--Howson algorithm, and support enumeration.
\end{abstract}

{\small

}
\vskip2ex
\section{Introduction}

A bimatrix game is a two-player game in strategic form.
The Nash equilibria of a bimatrix game correspond to pairs
of vertices of two polyhedra derived from the payoff
matrices.
These vertex pairs have to be ``completely labeled'',
which expresses the equilibrium condition that every pure
strategy of a player (represented by a ``label'') is either
a best response to the other player's mixed strategy or 
played with probability zero.

This polyhedral view gives rise to algorithms that compute a
Nash equilibrium.
A classical method is the algorithm by Lemke and
Howson (1964) which follows a path of ``almost completely
labeled'' polytope edges that terminates at Nash
equilibrium.
The Lemke--Howson (LH) algorithm has been one inspiration for the complexity
class PPAD defined by Papadimitriou (1994) of computational
problems defined by such path-following arguments, which
includes more general equilibrium problems such as the
computation of approximate Brouwer fixed points.
An important result proved by Chen and Deng (2006) and
Daskalakis, Goldberg, and Papadimitriou (2009) states that
every problem in the class PPAD can be reduced to finding a
Nash equilibrium of a bimatrix game, which makes this
problem ``PPAD-complete''.
(The problem of finding the Nash equilibrium at the end of a
\emph{specific} path is a much harder, namely
PSPACE-complete, see Goldberg, Papadimitriou, and Savani 2013.)

If an algorithm takes exponentially many steps (measured in
the size of its input) for certain problem instances, these
are considered ``hard'' instances for the algorithm.
Savani and von Stengel (2006) constructed bimatrix games
that are hard instances for the LH algorithm.
Their construction uses ``dual cyclic polytopes'' which
have a well-known vertex structure for any dimension and
number of linear inequalities.
Morris (1994) used similarly labeled dual cyclic polytopes
where all ``Lemke paths'' are exponentially long.
A Lemke path is related to the path computed by the LH
algorithm, but is defined on a \emph{single} polytope that
does not have a product structure corresponding to a
bimatrix game.
The completely labeled vertex found by a Lemke path can be
interpreted as a symmetric equilibrium of a symmetric bimatrix
game.
However, as in the example in Figure~\ref{fbr} below, such a 
symmetric game may also have nonsymmetric equilibria which
here are easy to compute, so that the result by Morris
(1994) seemed unsuitable to describe games that are hard
to solve with the LH algorithm.

The ``imitation games'' defined by McLennan and Tourky
(2010) changed this picture.
In an imitation game, the payoff matrix of one of the
players is the identity matrix.
The mixed strategy of that player in any Nash equilibrium of
the imitation game corresponds exactly to a symmetric
equilibrium of the symmetric game defined by the payoff
matrix of the other player.
In that way, an algorithm that finds a Nash equilibrium of a
bimatrix game can be used to find a symmetric Nash
equilibrium of a symmetric game.
(The converse statement that a bimatrix game can be
``symmetrized'', see Proposition~\ref{p-sym} below, is an
earlier folklore result stated for zero-sum games by Gale,
Kuhn, and Tucker 1950.)

In one sense the two-polytope construction of Savani and von
Stengel (2006) was overly complicated: the imitation games
by McLennan and Tourky (2010) provide a simple and elegant
way to turn the single-polytope construction of Morris
(1994) into exponentially-long LH paths for bimatrix games.
In another sense, the construction of Savani and von Stengel
was not redundant.
Namely, the square imitation games obtained from Morris
(1994) have a single completely mixed equilibrium that is
easily computed by equating all payoffs for all pure
strategies.
Savani and von Stengel (2006) extended their construction of
square games with long LH paths (and a single completely
mixed equilibrium) to non-square games that are
\emph{simultaneously} hard for the LH algorithm and
``support enumeration'', which is another natural and simple
algorithm for finding equilibria.
The support of a mixed strategy is the set of pure
strategies that are played with positive probability.
Given a pair of supports of equal size, the mixed strategy
probabilities are found by equating all payoffs for the
other player's support, which then have to be compared with
payoffs outside the support to establish the equilibrium
property (see Dickhaut and Kaplan 1991).

In this paper, we extend the idea of imitation games to
games where one payoff matrix is arbitrary and the other is
a set of unit vectors.
We call these \emph{unit vector games}.
An imitation game is an example of a unit vector game, where
the unit vectors form an identity matrix.
The main result of this paper is an application of
unit vector games: we use them to extend Morris's
construction to obtain non-square bimatrix games that use only one dual
cyclic polytope, rather than the two used by Savani and von
Stengel, and which are simultaneously hard \emph{both} for
the LH algorithm and support enumeration.
This result (Theorem~\ref{t-main}) was first described by
Savani (2006, Section~3.8).

Before presenting this construction in Section~\ref{s-hard}, we
introduce in Section~\ref{s-unit} the required background on labeled
best response polytopes for bimatrix games, in an accessible presentation due
to Shapley (1974) that we think every game theorist should
know.
We define unit vector games and the use of imitation games,
and their relationships to the LH algorithm.
We will make the case that unit vector games provide a
general and simple way to construct bimatrix games using a
single labeled polytope.

To our knowledge, unit vector games were first defined and
used by Balthasar (2009, Lemma 4.10) in a different context,
namely in order to prove that a symmetric equilibrium of a
nondegenerate symmetric game that has positive ``symmetric
index'' can be made the \emph{unique} symmetric equilibrium
of a larger symmetric game by adding suitable strategies
(Balthasar 2009, Theorem~4.1).

\section{Unit vector games}
\label{s-unit}

In this section, we first describe in Section~\ref{s-poly} how
labeled polyhedra capture the ``best-response regions'' of
mixed strategies where a particular pure strategy of the
other player is a best response, and how these are used
to identify Nash equilibria.
In Section~\ref{s-unitv} we introduce unit vector games, whose
equilibria correspond to completely labeled vertices of a single labeled
polytope.
In Section~\ref{s-redu} we discuss the role of imitation
games for symmetric games and their symmetric equilibria.
Finally, in Section~\ref{s-lemke}, we show how Lemke paths defined for
single labeled polytopes are ``projections'' of
seemingly more general LH paths in the case of unit vector
games (Theorem~\ref{t-proj}).

\subsection{Nash equilibria of bimatrix games and polytopes}
\label{s-poly}

Consider an $m\times n$ bimatrix game $(A,B)$.
We describe a geometric-combinatorial ``labeling'' method,
due to Shapley (1974), that allows an easy identification of
the Nash equilibria of the game.
It has an equivalent description in terms of polytopes
derived from the payoff matrices.

Let $\0$ be the all-zero vector and let $\1$ be the all-one
vector of appropriate dimension.
All vectors are column vectors and $C\T$ is the transpose of
any matrix $C$, so $\1\T$ is the all-one row vector.
Let $X$ and $Y$ be the mixed-strategy simplices of the two
players,
\begin{equation}
\label{XY}
X = \{x\in\reals^m\mid x\ge\0,~\1\T x=1\,\},
\qquad
Y = \{y\in\reals^n\mid y\ge\0,~\1\T y=1\,\}.
\end{equation}
It is convenient to identify the $m+n$ pure strategies of
the two players by separate \emph{labels} where the labels
$1,\ldots,m$ denote the $m$ pure strategies of the row
player~1 and the labels $m+1,\ldots,$ $m+n$ denote the $n$ pure
strategies of the column player~2.

Consider mixed strategies $x\in X$ and $y\in Y$.
We say that $x$ has label $m+j$ for $1\le j\le n$ if $j$ is a
pure best response of player~2 to~$x$.
Similarly, $y$ has label $i$ for $1\le i\le m$ if $i$ is a
pure best response of player~1 to~$y$.
In addition, we say that
$x$ has label $i$ for $1\le i\le m$ if $x_i=0$, and that 
$y$ has label $m+j$ for $1\le j\le n$ if $y_j=0$.
That is, a mixed strategy such as $x$ has label $i$ (one of
the player's own pure strategies) if $i$ is not played.

In a Nash equilibrium, every pure strategy that is played
with positive probability is a best response to the
other player's mixed strategy.
In other words, if a pure strategy is not a best response,
it is played with probability zero.
Hence, a mixed strategy pair $(x,y)$ is a Nash equilibrium
if and only if every label in $\{1,\ldots,m+n\}$ appears as
a label of $x$ or of~$y$.
The Nash equilibria are therefore exactly those pairs
$(x,y)$ in $X\times Y$ that are \emph{completely labeled}
in this sense.

\begin{figure}[hbt]
\strut\hfill
\includegraphics[width=27ex]{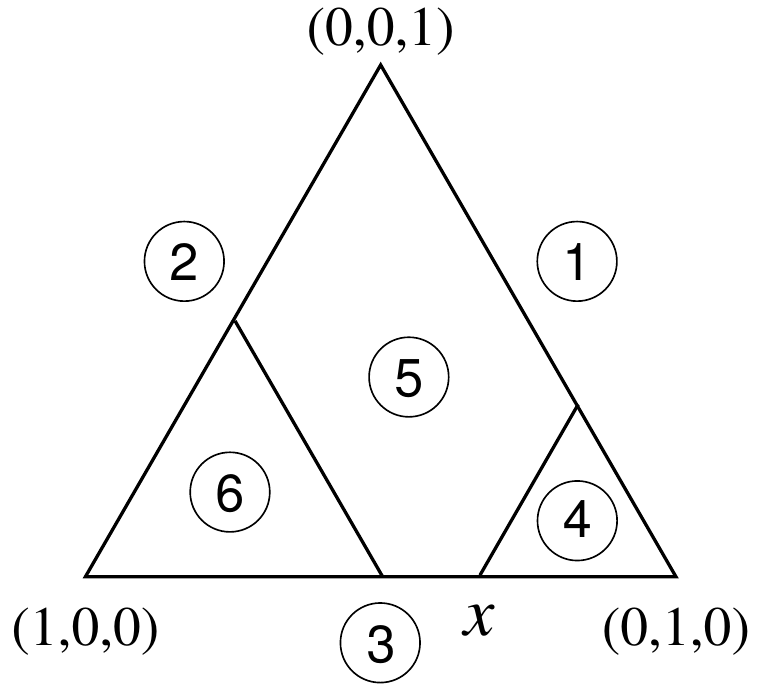}%
\hfill
%\vrule height 52mm ~ \vrule height 27ex
\hfill
\includegraphics[width=27ex]{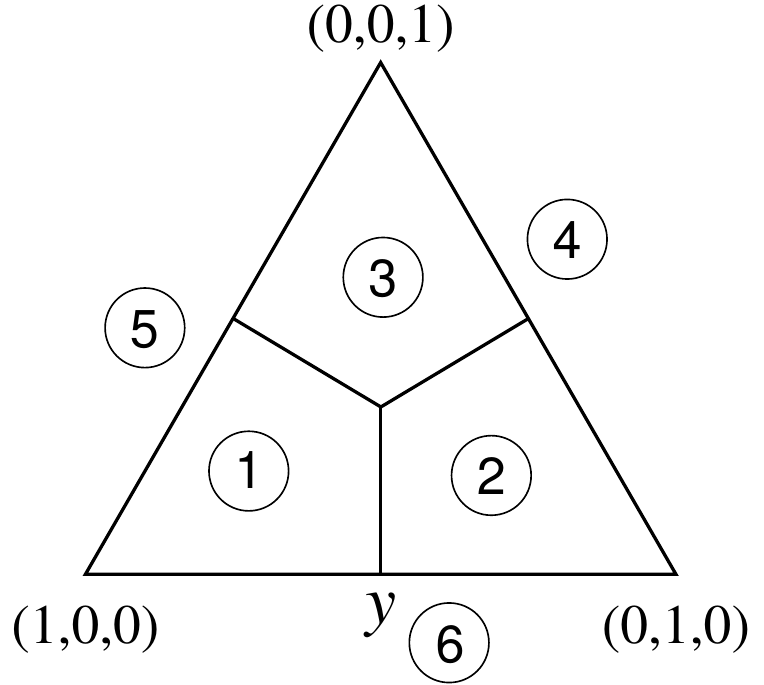}%
\hfill\strut
\caption{%
Labeled mixed strategy sets $X$ and $Y$ for the game
(\ref{AB}).
}
\label{f1}
\end{figure}

As an example, consider the $3\times 3$ game $(A,B)$ with
\begin{equation}
\label{AB}
A = \left(\begin{matrix}1&0&0\\ 0&1&0\\
0&0&1\end{matrix}\right),
\qquad
B = \left(\begin{matrix}0&2&4\\ 3&2&0\\
0&2&0\end{matrix}\right). 
\end{equation}
The labels $1,2,3$ represent the pure strategies of player~1
and $4,5,6$ those of player~2.
Figure~\ref{f1} shows $X$ and $Y$ with these labels shown as
circled numbers.
The interiors of these triangles are covered by
\emph{best-response regions} labeled by the other player's
pure strategies, which are closed polyhedral sets where the
respective pure strategy is a best response.
For example, the best-response region in $Y$ with label~$1$
is the set of those $(y_1,y_2,y_3)$ such that $y_1\ge y_2$ and
$y_1\ge y_3$, due to the particularly simple form of $A$ in
(\ref{AB}).  
The outsides of $X$ and $Y$ are labeled with the players'
own pure strategies where these are not played.
These outside facets are opposite to the vertex where only
that pure strategy is played; for example, label~$1$ is the
label of the facet of $X$ opposite to the vertex $(1,0,0)$.  
In Figure~\ref{f1} there is only one pair $(x,y)$
that is completely labeled, namely $x=(\frac13,\frac23,0)$
with labels $3,4,5$ and $y=(\frac12,\frac12,0)$ with labels
$1,2,6$, so this is the only Nash equilibrium of the game.

\begin{figure}[hbt]
\includegraphics[height=40mm]{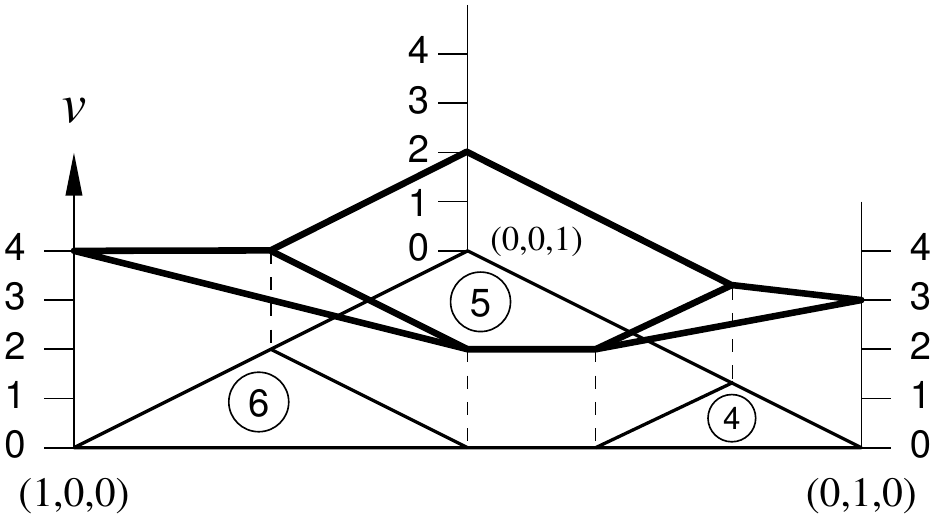}%
\hfill
\includegraphics[height=50mm]{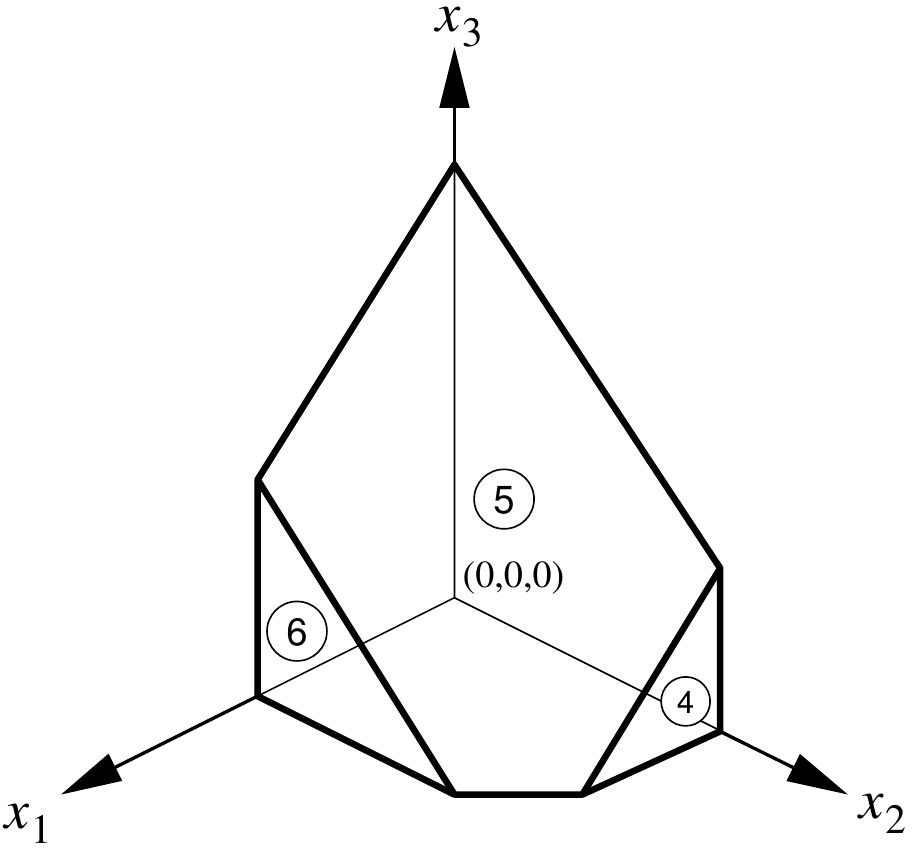}%
\caption{%
Best-response facets of the polyhedron $\overline P$ in
(\ref{PQbar}), and the polytope $P$ in (\ref{PQ}), for the
game in (\ref{AB}).
}
\label{fpol}
\end{figure}

The subdivision of $X$ and $Y$ into best-response regions is
most easily seen with the help of the ``upper envelope'' of the
payoffs to the other player, which are defined by the
following polyhedra.
Let
\begin{equation}
\label{PQbar}
\overline P= \{(x,v)\in X\times\reals\mid B\T x\le\1v\,\},
\qquad
\overline Q= \{(y,u)\in Y\times\reals\mid Ay\le\1u\,\}.
\end{equation}
For the example (\ref{AB}), the inequalities $B\T x\le\1v$
state that $3x_2\le v$, $2x_1+2x_2+2x_3\le v$, $4x_1\le v$, which
say that $v$ is at least the best-response payoff to
player~2.
If one of these inequalities is tight (holds as equality),
then $v$ is exactly the best-response payoff to player~2.
The left-hand diagram in Figure~\ref{fpol} shows these
``best-response facets'' of $\overline P$, and their
projection to $X$ by ignoring the payoff variable~$v$, which
defines the subdivision of $X$ into best-response regions as
in the left-hand diagram in Figure~\ref{f1}.  

Throughout this paper, assume (without loss of generality)
that $A$ and $B\T$ are non-negative and have no zero column.
Then $v$ and $u$ in $B\T x\le\1v$ and $Ay\le\1u$ are always
positive.
By dividing these inequalities by $v$ and $u$, respectively,
and writing $x_i$ instead of $x_i/v$ and $y_j$ instead of
$y_j/u$, the polyhedra $\overline P$ and $\overline Q$ are
replaced by $P$ and $Q$, 
\begin{equation}
\label{PQ}
P=\{x\in\reals^m\mid x\ge\0,~B\T x\le\1\},
\qquad
Q=\{y\in\reals^n\mid Ay\le\1,~ y\ge\0\},
\end{equation}
which are bounded and therefore polytopes.
For $B$ in (\ref{AB}), $P$ is shown on the right in
Figure~\ref{fpol}.

Both polytopes $P$ and $Q$ in (\ref{PQ}) are defined by
$m+n$ inequalities that correspond to the pure strategies of
the player, which we have denoted by the labels
$1,\ldots,m+n$.
We can now identify the labels, as pure best responses of
the other player, or unplayed own pure strategies, as tight
inequalities in either polytope.
That is, a point $x$ in $P$ has label $k$ if the $k$th
inequality in $P$ is tight, that is,
if $x_k=0$ for $1\le k\le m$
or $(B\T x)_{k-m}=1$ for $m+1\le k\le m+n$.
Similarly, $y$ in $Q$ has label $k$ if
$(Ay)_k=1$ for $1\le k\le m$
or $y_{k-m}=0$ for $m+1\le k\le m+n$.
Then $(x,y)$ in $P\times Q$ is \emph{completely labeled} if
$x$ and $y$ together have all labels in $\{1,\ldots,m+n\}$.
With the exception of $(\0,\0)$, these completely labeled
points of $P\times Q$ represent (after re\-scaling to become
pairs of mixed strategies) exactly the Nash equilibria of
the game $(A,B)$.

The pair $(x,y)$ in $P\times Q$ is completely labeled if 
\begin{equation}
\label{compl}
x_i=0 \hbox{ or }(Ay)_i=1 \hbox{ for all }i=1,\ldots,m,
\qquad
y_j=0 \hbox{ or }(B\T x)_j=1 \hbox{ for all }j=1,\ldots,n.
\end{equation}
Because %for $x\in P$ and $y\in Q$,
$x$, $\1-Ay$, $y$, and $\1-B\T x$ are all non-negative, the
\emph{complementarity} condition (\ref{compl}) can also be
stated as the orthogonality condition
\begin{equation}
\label{orth}
x\T(\1-Ay)=0,
\qquad
y\T(\1-B\T x)=0. 
\end{equation} 

The characterization of Nash equilibria as completely
labeled pairs $(x,y)$ holds for arbitrary bimatrix games.
For considering algorithms, it is useful to assume that the
game is \emph{non\-degen\-erate} in the sense that no point in
$P$ has more than $m$ labels, and no point in $Q$ has more
than $n$ labels.
Clearly, for a non\-degen\-erate game, in an equilibrium
$(x,y)$ each label appears exactly once either as a label of
$x$ or of~$y$.

Nondegeneracy is equivalent to the condition that the
number of pure best responses against a mixed strategy is never
larger than the size of the support of that mixed strategy.
It implies that $P$ is a \emph{simple} polytope in the sense
that no point of $P$ lies on more than $m$ \emph{facets},
and similarly that $Q$ is a simple polytope.
A facet is obtained by turning one of the inequalities
that define the polytope into an equality, provided that the
inequality is irredundant, that is, cannot be omitted
without changing the polytope.
A redundant inequality in the definition of $P$ and $Q$ may
also give rise to a degeneracy if it corresponds to a pure
strategy that is weakly (but not strictly) dominated by, or
payoff equivalent to, a mixture of other strategies.  
For a detailed discussion of degeneracy see von Stengel
(2002).

\subsection{Unit vector games and a single labeled polytope}
\label{s-unitv}

The components of the $k$th \emph{unit vector} $e_k$ are
$0$ except for the $k$th component, which is~$1$. 
In an $m\times n$ \emph{unit vector game} $(A,B)$, every
column of $A$ is a unit vector in $\reals^m$.
The matrix $B$ is arbitrary, and without loss of generality $B\T$ is 
non-negative and has no zero column.

In this subsection, we consider such a unit vector game
$(A,B)$.
Let the $j$th column of $A$ be the unit vector
$e_{\ell(j)}$, for $1\le j\le n$.
Then the sequence $\ell(1),\ldots,\ell(n)$ together with
the payoff matrix $B$ completely specifies the game.

For this game, the polytope $Q$ in (\ref{PQ}) has a very
special structure.
For $1\le i\le m$, let
\begin{equation}
\label{Ni}
N_i = \{\,j\mid \ell(j)=i,~1\le j\le n\}
\end{equation}
so that $N_i$ is the set of those columns $j$ whose best
response is row~$i$.
These sets $N_i$ are pairwise disjoint, and their union is
$\{1,\ldots,n\}$.
Then clearly
\begin{equation}
\label{QNi}
Q = \{y\in\reals^n\mid \sum_{j\in N_i}y_j\le 1,~1\le i\le
m,~ y\ge\0\}.
\end{equation}
That is, except for the order of inequalities, $Q$ is the
product of $m$ simplices of the form
%$\{z\in\reals^{|N_i|}\mid \sum_{j\in N_i}z_j\le 1,
$\{z\in\reals^{N_i}\mid \sum_{j\in N_i}z_j\le 1,
~ z\ge\0\}$,
for $1\le i\le m$.
If each $N_i$ is a singleton, then, by (\ref{Ni}), $A$ is a
permuted identity matrix, $n=m$, each simplex is the unit
interval, and $Q$ is the $n$-dimensional unit cube.

In any bimatrix game, the polytopes $P$ and $Q$ in
(\ref{PQ}) each have $m+n$ inequalities that correspond to
the pure strategies of the two players.
Turning the $k$th inequality into an equality typically
defines a facet of the polytope, which defines the label $k$
of that facet, $1\le k\le m+n$.

In our unit vector game $(A,B)$ where the $j$th column of $A$
is the unit vector $e_{\ell(j)}$, for $1\le j\le n$, we
% can dispense with the polytope $Q$ altogether by
introduce the \emph{labeled polytope} $\Pell$, 
\begin{equation}
\label{Pell}
\Pell=\{x\in\reals^m\mid x\ge\0,~B\T x\le \1\},
\end{equation} 
where the $m+n$ inequalities of $\Pell$ have the labels $i$
for the first $m$ inequalities $x_i\ge0$, $1\le i\le m$, and
the $j$th inequality of $B\T x\le \1$ has label $\ell(j)$,
for $1\le j\le n$.
That is, $\Pell$ is just the polytope $P$ in (\ref{PQ})
except that the last $n$ inequalities are labeled with
$\ell(1),\ldots,\ell(n)$, each of which is a number in
$\{1,\ldots,m\}$.
A point $x$ of $\Pell$ is \emph{completely labeled} if
every number in $\{1,\ldots,m\}$ appears as the 
label of an inequality that is tight for~$x$.
In particular, if $\Pell$ is a simple polytope with one label
for each facet, then $x$ is completely labeled if $x$ is a
vertex of $\Pell$ so that the $m$ facets that $x$ lies on
together have all labels $1,\ldots,m$.

The following proposition shows that with these labels,
$\Pell$ carries all the information about the unit vector
game, and the polytope $Q$ is not needed.
The proposition was first stated in a dual version
by Balthasar (2009, Lemma 4.10), and in essentially this
form by V\'egh and von Stengel (2015, Proposition~1).
Its proof also provides the first step of the proof of
Theorem~\ref{t-proj} below.
% Vegh von Stengel p18: Todd simplicial duoid

\begin{proposition}
\label{p-unitv}
Consider a labeled polytope $\Pell$ with labels as
described following $(\ref{Pell})$.
Then $x$ is a completely labeled point of $\Pell-\{\0\}$ if and
only if for some~$y\in Q$ the pair $(x,y)$ is (after scaling)
a Nash equilibrium of the $m\times n$ unit vector game
$(A,B)$ where $A=[e_{\ell(1)}\cdots e_{\ell(n)}]$.
\end{proposition}

\proof
Let $(x,y)\in P\times Q-\{(\0,\0)\}$ be a Nash equilibrium,
so it has all labels in $\{1,\ldots,m+n\}$.
Then $x$ is a completely labeled point of $\Pell$ for the
following reason.
If $x_i=0$ then $x$ has label~$i$.
If $x_i>0$ then $y$ has label~$i$, that is, $(Ay)_i=1$,
which requires that for some $j$ we have $y_j>0$ and the
$j$th column of $A$ is equal to $e_i$, that is, $\ell(j)=i$.
Because $y_j>0$, and $(x,y)$ is completely labeled, $x$ has
label $m+j$ in $P$, that is, $(B\T x)_j=1$, which means $x$
has label $\ell(j)=i$ in $\Pell$, as required.

Conversely, let $x$ be a completely labeled point of
$\Pell-\{\0\}$.
Then for each $i$ in $\{1,\ldots,m\}$ with $x_i>0$,
label $i$ for $x$ comes from a binding inequality
$(B\T x)_j=1$ with label $\ell(j)=i$, that is, for some
$j\in N_i$ in (\ref{Ni}).
Let $y_j=1$ and $y_h=0$ for all $h\in N_i-\{j\}$, and
do this for all $i$ with $x_i>0$.
It is easy to see that the pair $(x,y)$ is a completely
labeled point of $P\times Q$.
\endproof

\begin{figure}[hbt]
\strut\hfill
\includegraphics[width=27ex]{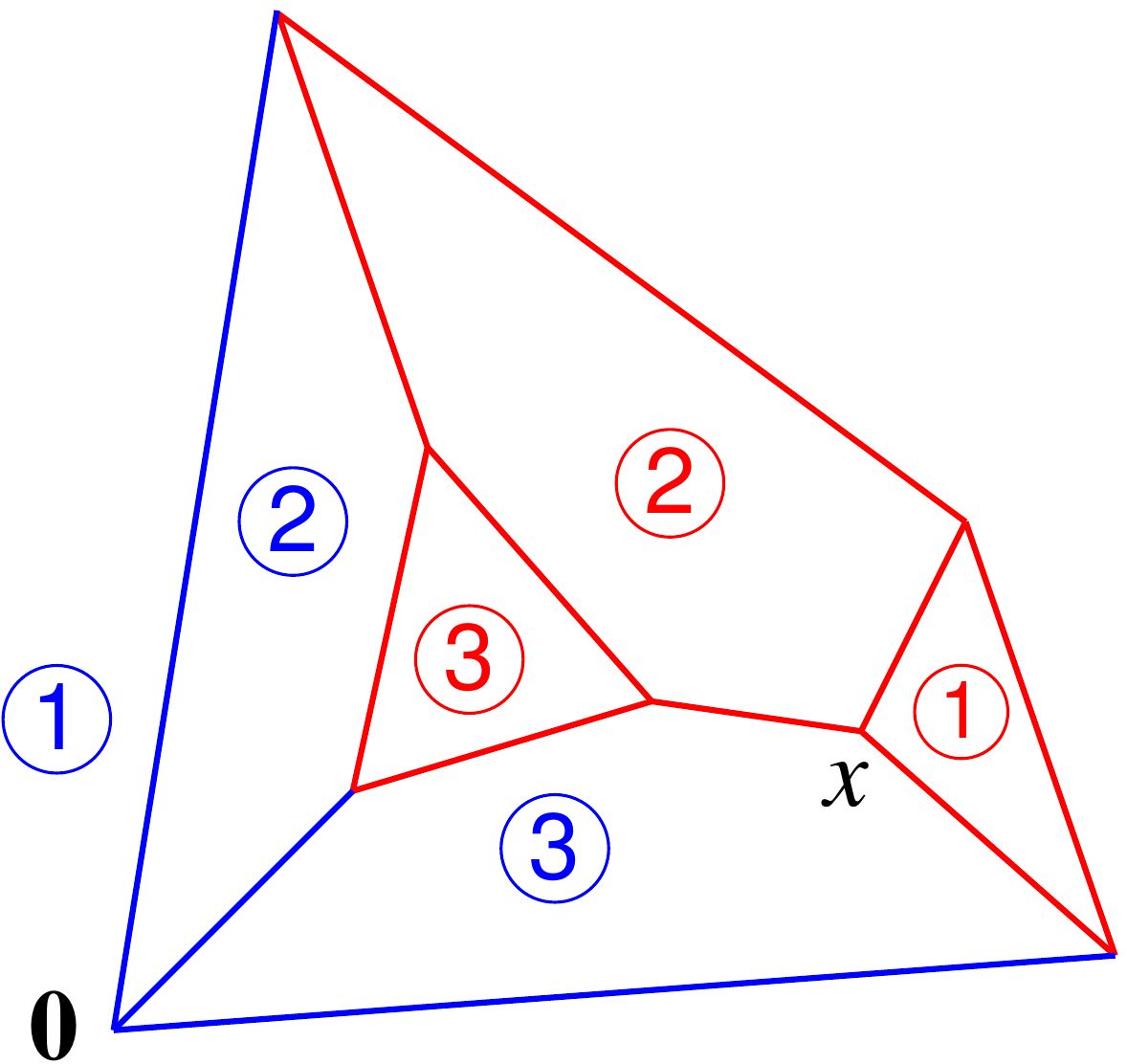}%
\hfill\strut
\caption{% 
The polytope $\Pell$ for the unit vector game (\ref{AB}).
The hidden facet at the back has label~$1$, written on the
left.
}
\label{fpell}
\end{figure}

The game in (\ref{AB}) is a unit vector game.
For this game, the polytope $P$ in (\ref{PQ}) is shown on
the right in Figure~\ref{fpol}, where we have shown only the
labels $4,5,6$ for the ``best response facets''.
In addition, the  facets with labels $1,2,3$ where $x_1=0$,
$x_2=0$, $x_3=0$ are the facets, hidden in this picture,
at the back right, back left, and bottom of the polytope,
respectively.
In the polytope $\Pell$, the labels $4,5,6$ are replaced by
$1,2,3$ because the corresponding columns of $A$ are the
unit vectors $e_1,e_2,e_3$.
Figure~\ref{fpell} shows this polytope in such a way that
there is only one hidden facet, with label~$1$ where
$x_1=0$.
Apart from the origin $\0$, the only completely labeled
point of $\Pell$ is $x$ as shown, which is part of a Nash
equilibrium $(x,y)$ as stated in Proposition~\ref{p-unitv}.

A polytope like in (\ref{Pell}) that has a label for each
facet provides a particularly natural way to describe
equilibrium-finding algorithms, as described in
Section~\ref{s-lemke} below.
% At the same time, unit vector games are completely general,
% as we see next.

\subsection{Reductions between equilibria of 
bimatrix and unit vector games}
\label{s-redu}
%and symmetric equilibria of symmetric games}

A method that ``solves'' a unit vector game in the sense of
finding one equilibrium, or all equilibria, of the game, can
be used to solve an arbitrary bimatrix game.
The first step in seeing this is the fact that the equilibria
of a bimatrix game correspond to the symmetric equilibria of
a suitable symmetric game.
This ``symmetrization'' has been observed for zero-sum games
by Gale, Kuhn, and Tucker (1950) and seems to be a folklore
result for bimatrix games.

\begin{proposition}
\label{p-sym}
Let $(A,B)$ be a bimatrix game, and $(x,y)\in P\times
Q-\{(\0,\0)\}$ in $(\ref{PQ})$.
Then $(x,y)$ (suitably scaled) is a Nash equilibrium of $(A,B)$
if and only if $(z,z)$ (suitably scaled)
is a symmetric equilibrium of $(C,C\T)$ with
%$z=(x,y)$ and $C={~0~~~A \choose \,B\T\,~0\,}$.
$z=(x,y)$ and $C=\binom{~0~~~A}{\,B\T\,~0\,}$.
\end{proposition} 

\proof
This holds by (\ref{orth}) because $(z,z)$ is an equilibrium
of $(C,C\T)$ if and only if $z\ne\0$, $z\ge\0$, $Cz\le\1$,
and $z\T(\1-Cz)=0$.
\endproof

By Proposition~\ref{p-sym}, finding an equilibrium of a
bimatrix game can be reduced to finding a symmetric
equilibrium of a symmetric bimatrix game.
The converse follows from the following proposition, due to
McLennan and Tourky (2010, Proposition~2.1), with the help of
imitation games.
They define an imitation game as an $m\times m$ bimatrix
game $(A,B)$ where $B$ is the identity matrix.
Here, we define an imitation game as a special unit vector
game $(A,B)$ where $A$ (rather than~$B$) is the identity
matrix~$I$.
The reason for this (clearly not very material) change is
that this game is completely described by the polytope
$\Pell$ in (\ref{Pell}), which corresponds to $P$ in
(\ref{PQ}) and compared to $Q$ has a more natural
description because the $m$ inequalities $x\ge\0$ with
labels $1,\ldots,m$ are listed first.

\begin{proposition}
\label{p-imit}
The pair $(x,x)$ is a symmetric Nash equilibrium of the
symmetric bimatrix game $(C,C\T)$ if and only if $(x,y)$ is
a Nash equilibrium of the imitation game $(I,C\T)$ for
some~$y$.
\end{proposition}

\begin{figure}[hbt]
\strut\hfill
\includegraphics[width=27ex]{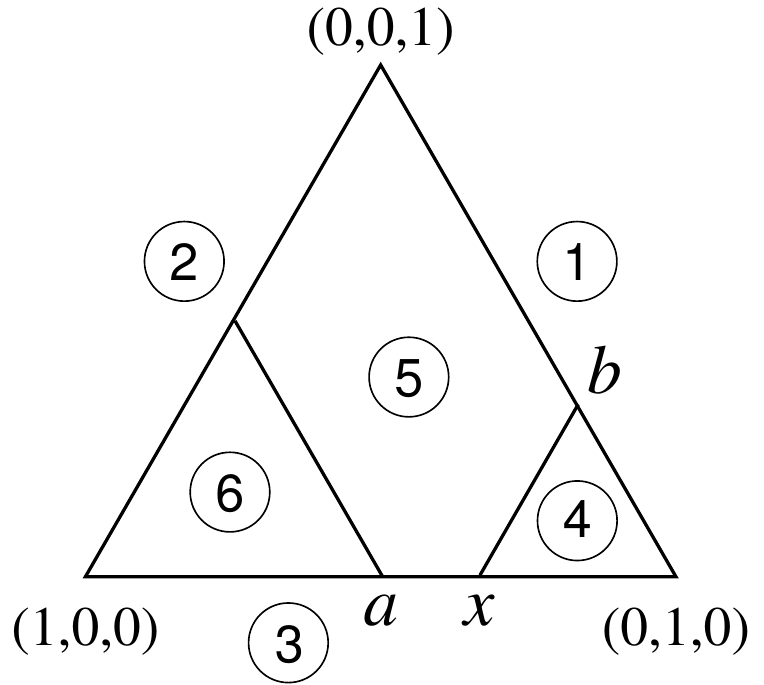}%
\hfill
\hfill
\includegraphics[width=27ex]{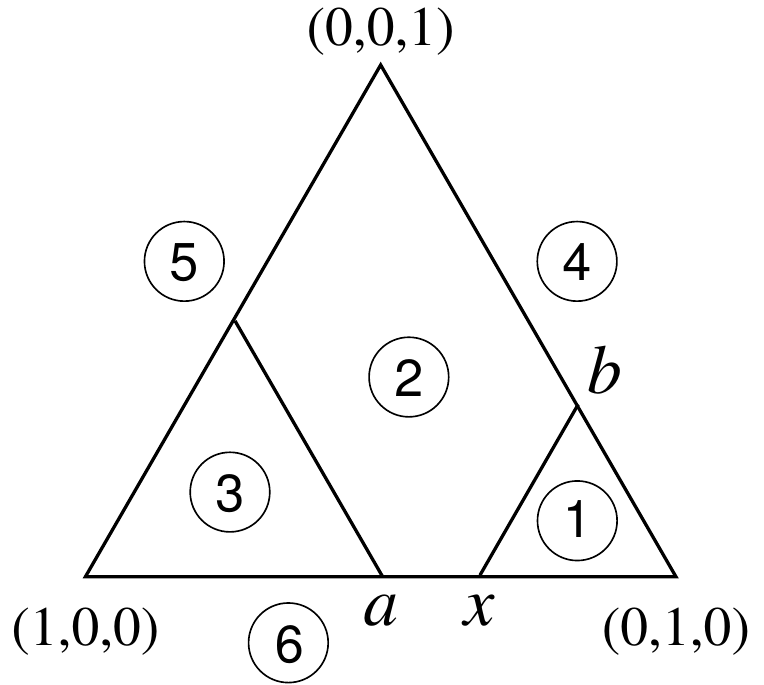}%
\hfill\strut
\caption{% 
Labeled mixed-strategy sets $X$ and $Y$ for the symmetric
game $(C,C\T)$ in (\ref{C}).
}
\label{fbr}
\end{figure}

% Example:      030
%           C = 222
%               400
As an example, consider the symmetric game $(C,C\T)$ with
\begin{equation}
\label{C}
C = \left(\begin{matrix}0&3&0\\ 2&2&2\\
4&0&0\end{matrix}\right),
\qquad
C\T = \left(\begin{matrix}0&2&4\\ 3&2&0\\
0&2&0\end{matrix}\right), 
\end{equation}
so that $C\T=B$ in (\ref{AB}).
Figure~\ref{fbr} shows the labeled mixed-strategy simplices
$X$ and $Y$ for this game.
In addition to the symmetric equilibrium $(x,x)$ where
$x=(\frac13,\frac23,0)$, the game has two non-symmetric
equilibria $(a,b)$ and $(b,a)$ where $a=(\frac12,\frac12,0)$
and $b=(0,\frac23,\frac13)$.
A method that just finds a Nash equilibrium of a bimatrix
game may not find a symmetric equilibrium when applied to
this game, which shows the use of Proposition~\ref{p-imit}.
The corresponding imitation game $(I,C\T)$ is just $(A,B)$
in (\ref{AB}), which has the unique equilibrium $(x,y)$
where $(x,x)$ is the symmetric equilibrium of $(C,C\T)$.

\begin{figure}[hbt]
\strut\hfill
\includegraphics[width=27ex]{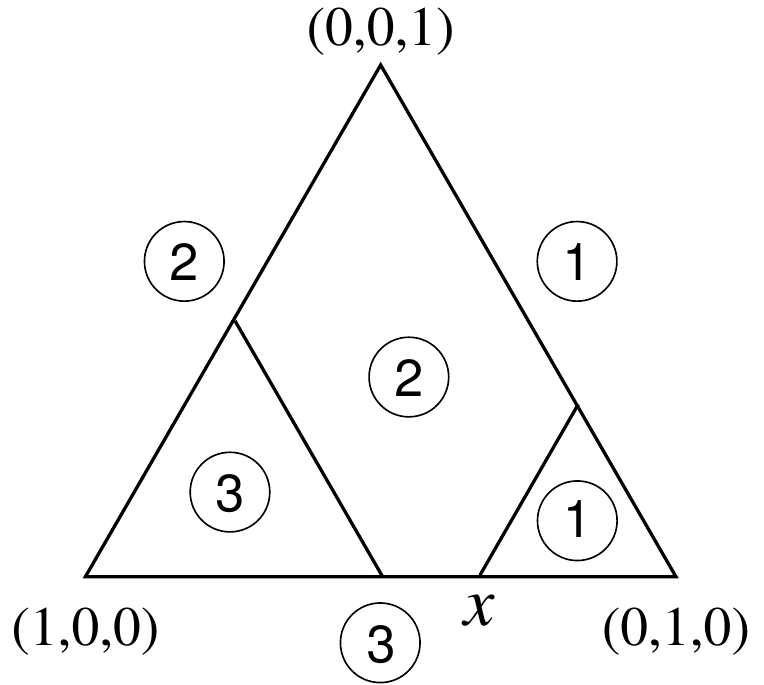}%
\hfill
\hfill
\includegraphics[width=27ex]{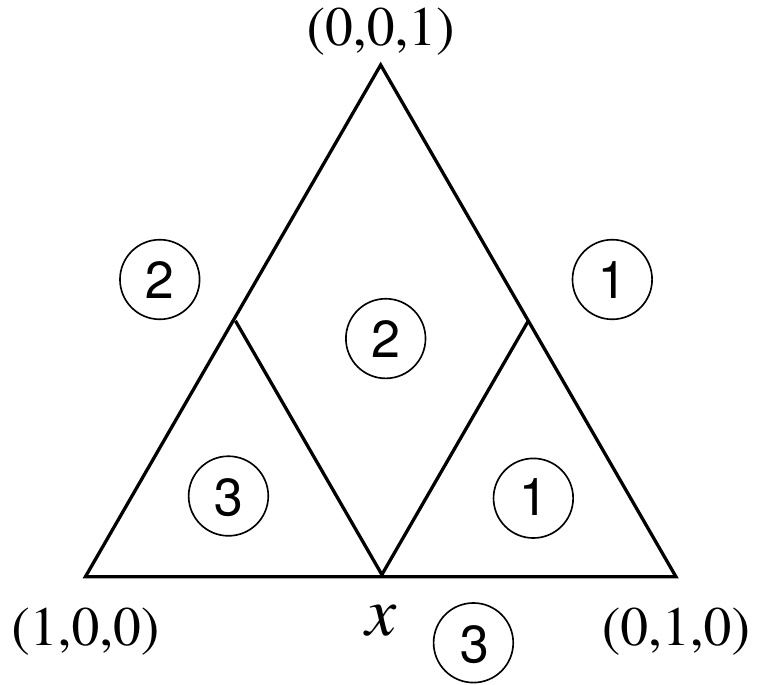}%
\hfill\strut
\caption{% 
(Left) Best-response regions for identifying symmetric
equilibria.
(Right) Degenerate symmetric game (\ref{Cdeg}) with a unique
symmetric equilibrium.
}
\label{fsingle}
\end{figure}

The left-hand diagram in Figure~\ref{fsingle} shows the mixed
strategy simplex $X$ subdivided into regions of pure best
responses against the mixed strategy itself, which corresponds to the
polytope $\Pell$ in Figure~\ref{fpell}.
The (in this case unique) symmetric equilibrium is the 
completely labeled point~$x$.

\begin{figure}[hbt]
\strut\hfill
\includegraphics[width=27ex]{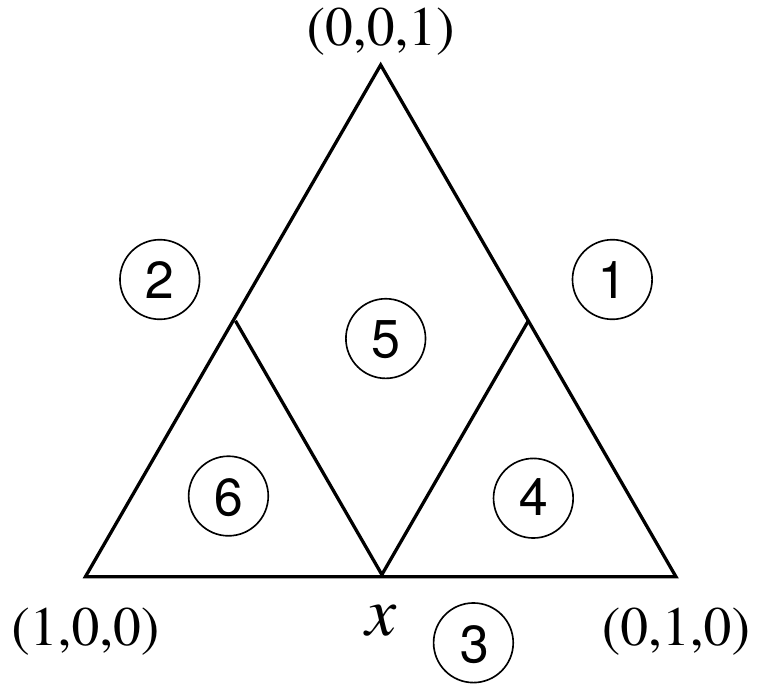}%
\hfill
\hfill
\includegraphics[width=27ex]{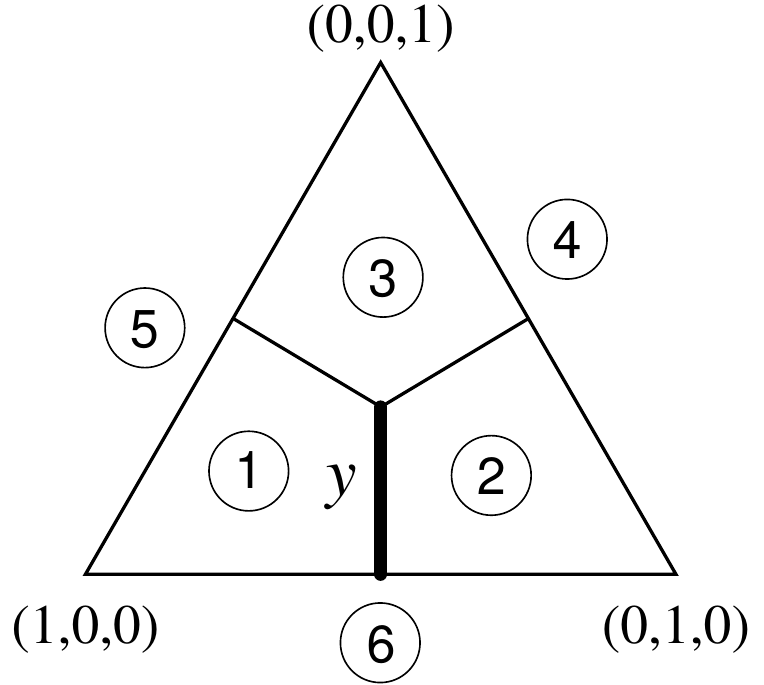}%
\hfill\strut
\caption{% 
Labeled mixed-strategy sets for the imitation game $(I,C\T)$ 
for the degenerate symmetric game (\ref{Cdeg}) where the
equilibria $(x,y)$ are not unique.
}
\label{fdeg}
\end{figure}

The right-hand diagram in Figure~\ref{fsingle} shows this
subdivision of $X$ for another game $(C,C\T)$ where
\begin{equation}
\label{Cdeg}
C = \left(\begin{matrix}0&4&0\\ 2&2&2\\
4&0&0\end{matrix}\right),
\qquad
C\T = \left(\begin{matrix}0&2&4\\ 4&2&0\\
0&2&0\end{matrix}\right). 
\end{equation}
This game is degenerate because the mixed strategy
$x=(\frac12,\frac12,0)$ has three pure best responses.
This mixed strategy $x$ also defines the unique symmetric equilibrium
$(x,x)$ of this game.
However, the corresponding equilibria $(x,y)$ of the
imitation game $(I,C\T)$ are not unique, because due to the
degeneracy any convex combination of $(\frac12,\frac12,0)$
and $(\frac13,\frac13,\frac13)$ can be chosen for~$y$, as
shown in Figure~\ref{fdeg}.

Hence the reduction between symmetric equilibria $(x,x)$ of
a symmetric game and Nash equilibria $(x,y)$ of the
corresponding imitation game stated in
Proposition~\ref{p-imit} does not preserve uniqueness if
the game is degenerate.

\subsection{Lemke paths and Lemke--Howson paths}
\label{s-lemke}

Consider a labeled polytope $\Pell$ as in (\ref{Pell}).
We assume throughout that $\Pell$ is nondegenerate, that is,
no point of $\Pell$ has more than $m$ labels. Therefore,
$\Pell$ is a simple polytope, and every tight inequality
defines a separate facet (we can omit inequalities that are never
tight), each of which has a label in $\{1,\ldots,m\}$.
The path-following methods described in this section can
be extended to degenerate games and polytopes; for an
exposition see von Stengel (2002).

\begin{figure}[hbt]
\strut\hfill
\includegraphics[width=27ex]{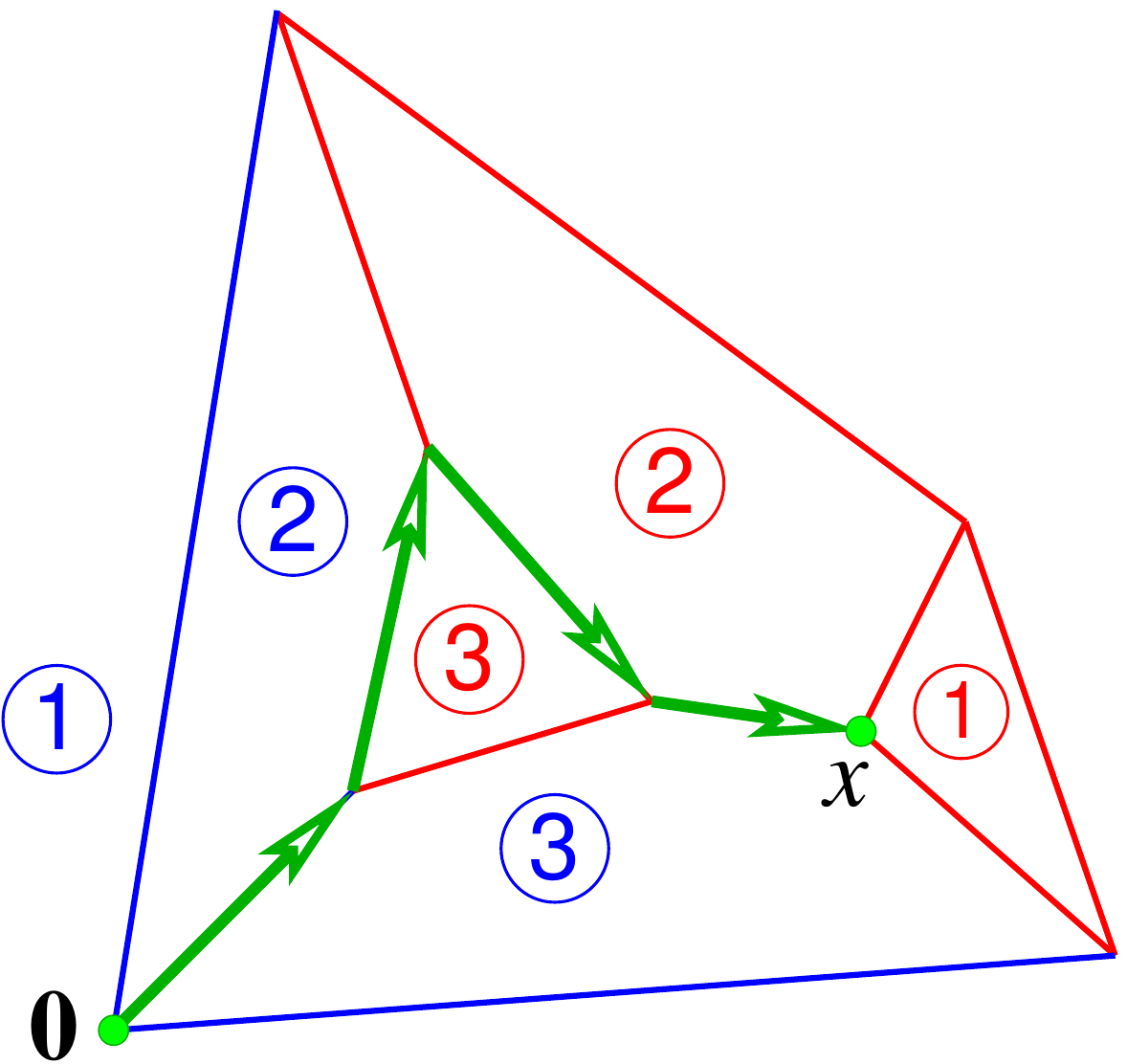}%
\hfill\strut
\caption{% 
Lemke path for missing label $1$ for the polytope in
Figure~\ref{fpell}.
}
\label{fpath}
\end{figure}

A \emph{Lemke path} is a path that starts at a
completely labeled vertex of $\Pell$ such as $\0$ and ends
at another completely labeled vertex.
It is defined by choosing one label $k$ in $\{1,\ldots,m\}$
that is allowed to be \emph{missing}.
After this choice of $k$, the path proceeds in a unique
manner from the starting point.
By leaving the facet with label $k$, a unique edge is
traversed whose endpoint is another vertex, which lies on a
new facet.
The label, say $j$, of that facet, is said to be
\emph{picked up}.
If this is the missing label $k$, then the path terminates
at a completely labeled vertex.
Otherwise, $j$ is clearly \emph{duplicate} and the next edge
is uniquely chosen by leaving the facet that so far had
label $j$, and the process is repeated.
The resulting path consists of a sequence of
\emph{$k$-almost complementary} edges and vertices (so
defined by having all labels except possibly $k$, where $k$
occurs only at the starting point and endpoint of the path).
The path cannot revisit a vertex because this would offer
a second way to proceed when that vertex is first
encountered, which is not the case because $\Pell$ is
nondegenerate.
Hence, the path terminates at another completely labeled
vertex of $\Pell$ (which is a Nash equilibrium of the
corresponding unit vector game in Proposition~\ref{p-unitv}
if the path starts at $\0$).
Figure~\ref{fpath} shows an example.

For a fixed missing label~$k$, every completely labeled
vertex of $\Pell$ is a separate endpoint of a Lemke path.
Because each path has two endpoints, there is an even number of them,
and all of these except $\0$ are Nash equilibria of the unit
vector game, so the number of Nash equilibria is odd.

This path-following method was first described by Lemke
(1965) in order to find a solution to a linear
complementarity problem (LCP);
it is normally described for polyhedra, not for polytopes,
so that termination requires additional assumptions (see
Cottle, Pang, and Stone 1992).
The standard description of an LCP assumes a square matrix
$B$ with labels $\ell(j)=j$ for $j=1,\ldots,m$.
Allowing $\Pell$ to have $m+n$ rather than $2m$ facets with
individual labels $\ell(j)$ for the last $n$ facets
corresponds to a \emph{generalized LCP} (sometimes also
called ``vertical LCP''), as studied in Cottle and Dantzig
(1970).
The term ``Lemke paths'' for polytopes is due to Morris
(1994).

The algorithm by Lemke and Howson (1964) finds one Nash
equilibrium of an $m\times n$ bi\-matrix game $(A,B)$.
Let $C=\binom{~0~~~A}{\,B\T\,~0\,}$ as in
Proposition~\ref{p-sym}.
Then one way to define a Lemke--Howson (LH) path for missing
label $k$ in $\{1,\ldots,m+n\}$ is as a Lemke path for
missing label $k$ for the labeled polytope
\begin{equation}
\label{Rell}
\Rell=\{z\in \reals^{m+n}\mid z\ge\0,~ Cz\le \1\,\}
\end{equation}
where the $2(m+n)$ inequalities of $\Rell$ have labels
$1,\ldots,m+n,1,\ldots,m+n$ (that is, $\ell(i)=i$ for
$i=1,\ldots,m+n$).

The more conventional way to define the LH algorithm is to
consider $P\times Q$ with $P$ and $Q$ as in (\ref{PQ}).
Clearly, with $z=(x,y)$, $\Rell$ in (\ref{Rell}) is equal to
$P\times Q$.
Starting from $(\0,\0)$, the chosen missing label $k$ is
a pure strategy of player~1 (for $1\le k\le m$) or
of player~2 (for $m+1\le k\le m+n$).
Instead of a single point $z$ that moves on the graph (of
vertices and edges) of $\Rell$, the pair $(x,y)$ (which
equals $z$) moves on $P\times Q$ by alternately moving $x$
on the graph of $P$ and $y$ on the graph of $Q$.
This alternate move of a pair of ``tokens'' can be nicely
shown for $3\times 3$ games on the two mixed strategy sets
$X$ and $Y$ subdivided into best-response regions as in
Figure~\ref{f1}, extended with the origin $\0$ (as done by
Shapley 1974).
This is obviously more accessible than a path on a
six-dimensional polytope, but requires keeping track of
the alternating tokens.

\begin{figure}[hbt]
\strut\hfill
\includegraphics[height=28ex]{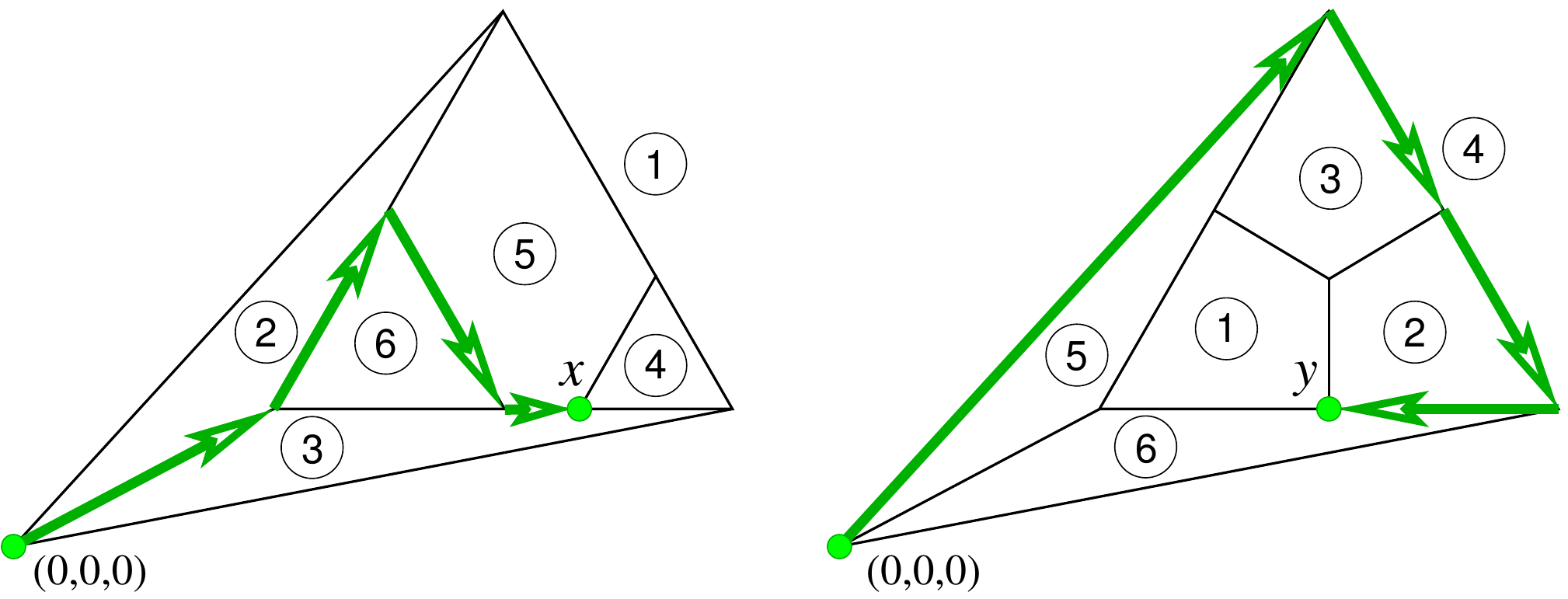}%
\hfill\strut
\caption{% 
Lemke--Howson path for missing label $1$ for the game (\ref{AB}).
}
\label{flh}
\end{figure}

Figure~\ref{flh} illustrates this for the game in
(\ref{AB}).
The pair of tokens starts on $\0,\0$, which is identified
by the pair of label sets $123,456$.
Let $1$ be the missing label, which means moving (in the
left-hand diagram in Figure~\ref{flh}) from $(0,0,0)$ with labels
$123$ to the vertex $(1,0,0)$ of $X$ with labels $236$.
The new pair has labels $236,456$ with duplicate label $6$,
so the next move is in the right-hand diagram from $(0,0,0)$ with
labels $456$ to the vertex $(0,0,1)$ of $Y$ with labels $345$.
The label that is picked up is $3$ which is now duplicate,
so the next move is in $X$ from $236$ to $256$.
Then $5$ is duplicate, with a move in $Y$ from $345$ to $234$.
With $2$ duplicate, the next move in $X$ is from $256$ to
$356$.
Then $3$ is duplicate, moving in $Y$ from $234$ to $246$.
Then $6$ is duplicate, moving in $X$ from $356$ to $345$,
which is the point $x=(\frac13,\frac23,0)$.
Then $4$ is duplicate, moving in $Y$ from $246$ to $126$,
which is the point $y=(\frac12,\frac12,0)$ which has the
missing label~$1$.
This terminates the LH path for missing label~$1$ at the
Nash equilibrium $(x,y)$.

The two diagrams in Figure~\ref{flh} show two separate
paths on $P$ and $Q$, respectively (represented by $X$ and
$Y$ subdivided into best-response regions).
These paths are traversed in alternate steps and define a
single path on the product polytope $P\times Q$.
In general, a simple path on $P\times Q$ (that is, a path
that does not revisit a vertex) may not ``project'' to 
simple paths on $P$ and~$Q$.
However, for LH paths this is the case, as stated in the
following proposition (Lemma 2.3 of Savani 2006, and
implicit in McLennan and Tourky 2010, Section~4).

\begin{proposition}
% Lemma 2.3 in thesis
\label{p-proj}
Every LH path on $P\times Q$ induces a simple path in
each polytope $P$ and $Q$, that is, no vertex of $P$ or $Q$
is ever left and visited again on an LH path.
\end{proposition}

\proof
Suppose to the contrary that a vertex $x$ of $P$ is left and
visited again on an LH path.
This means that there are three vertex pairs $(x,y)$,
$(x,y')$, and $(x,y'')$ of $P\times Q$, with pairwise distinct 
vertices $y$, $y'$, and $y''$ of $Q$, on an LH path with
missing label $k$, say.
All three pairs have all labels except possibly~$k$.
The $m$ labels of $x$ define $n-1$ labels shared by $y$,
$y'$, and $y''$.
However, this is impossible, since these $n-1$ labels
correspond to $n-1$ equations in $\reals^n$ that define a line,
which can only contain two vertices of~$Q$.
The same reasoning applies to a vertex $y$ of $Q$ that would
be visited multiple times on an LH path.
\endproof

Consider an $m\times n$ unit vector game $(A,B)$ where
$A=[e_{\ell(1)}\cdots e_{\ell(n)}]$.
According to Proposition~\ref{p-unitv}, the labeled polytope
$\Pell$ carries all information about the Nash equilibria of
$(A,B)$.
Recall that $\Pell$ is the polytope $P$ in (\ref{PQ}) but
where the labels $m+j$ for the strategies $j$ of the column
player, $1\le j\le n$, are replaced by $\ell(j)$, that is,
by the best responses of the row player to these columns.
Replacing these labels in the left-hand diagram in Figure~\ref{f1}
gives the left-hand diagram in Figure~\ref{fsingle}, equivalent
to $\Pell$ in Figure~\ref{fpell},

We now establish the same correspondence with regard to the
LH paths on $P\times Q$ for the game $(A,B)$, where the
corresponding ``projection'' to $P$ defines a Lemke path on
$\Pell$.
For example, the LH path projected to $P$ shown in the
left-hand diagram in Figure~\ref{flh} is the same as the Lemke
path on $\Pell$ in Figure~\ref{fpath}.
Both paths are defined for the missing label~$1$.
It seems natural that the LH path for missing label $i$ in
$\{1,\ldots,m\}$ projects to the Lemke path for missing 
label~$i$ on $\Pell$.
However, there are $n$ additional LH paths for the game
$(A,B)$ for the missing labels $m+j$ for
$j$ in $\{1,\ldots,n\}$, which do not exist as labels of
$\Pell$. 
The following theorem states that these project to the
Lemke paths on $\Pell$ for the missing label $\ell(j)$.
This generalizes the corresponding assertion by McLennan and
Tourky (2010, p.~9) and Savani and von Stengel (2006,
Proposition~15) for imitation games where $\ell(j)=j$.

\begin{theorem}
\label{t-proj}
Consider an $m\times n$ unit vector game $(A,B)$ where
$A=[e_{\ell(1)}\cdots e_{\ell(n)}]$, with $\Pell$ as in
$(\ref{Pell})$ and $P$ and $Q$ as in $(\ref{PQ})$.
Then the LH path on $P\times Q$ for this game for missing
label $k$ projects to a path on $P$ that is the Lemke path
on $\Pell$ for missing label $k$ if $1\le k\le m$, and that
is the Lemke path for missing label $\ell(j)$ if $k=m+j$ for
$1\le j\le n$.
\end{theorem}

\proof
In Proposition~\ref{p-unitv} it was shown that the
completely labeled pairs $(x,y)$ of $P\times Q$
cor\-res\-pond to the completely labeled points $x$ of $\Pell$.
It is easy to see that if $P$ is nondegenerate, as assumed
here, then this correspondence is one-to-one, and $x$ and
$y$ are vertices.

In the following,
$i$ is always an element of $\{1,\ldots,m\}$,
and $j$ is always an element of $\{1,\ldots,n\}$.

Consider a step of an LH path on $P\times Q$ that leaves
or arrives at a vertex $x$ of $P$, as part of a pair
$(x,y)$.
If the dropped label is $i$, 
then $x_i=0$ changes to $x_i>0$, and 
if the dropped label is $m+j$, then
$(B\T x)_j=1$ changes to $(B\T x)_j<1$.
If $i$ is a label that is picked up, then
$x_i>0$ changes to $x_i=0$, and 
if $m+j$ is a label that is picked up, then
$(B\T x)_j<1$ changes to $(B\T x)_j=1$.

Similarly, consider a vertex $y$ of $Q$.
Because $Q$ is a product of $m$ simplices as in (\ref{QNi}),
for each $i$ the following holds:
either $y_j=0$ for all $j\in N_i$, or for exactly one
$j\in N_i$ we have $y_j=1$ (which means $(Ay)_i=1$ and $y$
has label~$\ell(j)=i$) and $y_h=0$ for all $h\in N_i-\{j\}$.
We can also describe precisely which label is picked up
after moving away from $y$ by dropping a label:
% on an LH path on $P\times Q$:
\abs{(a)}
If the dropped label is $i$, 
then $y_j=1$ (for some $j\in N_i$) changes to $y_j=0$,
so that $m+j$ is the label that is picked up.
\abs{(b)}
If the dropped label is $m+j$, this is just the reverse step:
$j\in N_i$ for a unique $i=\ell(j)$, so
$y_j=0$ changes to $y_j=1$, which means $i=\ell(j)$ is the
label that is picked up.

Consider now steps on an LH path with missing label~$k$,
and assume that any label that is picked up is not the
missing label~$k$, and therefore duplicate.
Suppose label $i$ is picked up in $P$, corresponding to the 
binding inequality $x_i=0$.
Label $i$ is duplicate and therefore dropped in $Q$.
By (a), this means that $m+j$ with $j\in N_i$ is picked up
in $Q$, where $i=\ell(j)$.
The duplicate label $m+j$ in $P$ corresponds to the binding
inequality $(B\T x)_j=1$.
So the next step is to move away from this facet in $P$.
In $\Pell$ this same facet with $(B\T x)_j=1$ has label
$\ell(j)=i$, and moving away from this facet is exactly the
next step on the Lemke path on $\Pell$.

Similarly, suppose label $m+j$ is picked up in $P$, which
corresponds to the facet $(B\T x)_j=1$ which in $\Pell$ has
label $i=\ell(j)$.
On the LH path, the duplicate label $m+j$ in $Q$ is dropped
as in (b), where label $i$ is picked up in $Q$ and therefore
duplicate.
In $P$, the facet with this duplicate label is given by
$x_i=0$.
The next step on the LH path is to move away from this
facet, which is the same facet from which the Lemke path on
$\Pell$ moves away.

Similar considerations apply when the LH path is started or
terminates.
If the missing label is $k$ in $\{1,\ldots,m\}$, then
the LH path starts by dropping $k$ in $P$, and the Lemke
path starts in the same way in $\Pell$.
When the LH path terminates by picking up the missing label
$k$ in $P$, the Lemke path ends in the same way in $\Pell$.
If it terminates by picking up the missing label $k$ in $Q$,
then by (b) this was preceded by dropping the previously
duplicate label $m+j$ where $j\in N_k$, that is, after the
path reached in $P$ the facet defined by $(B\T x)_j=1$ which
has label $\ell(j)=k$ in $\Pell$, so the Lemke path has
already terminated on $\Pell$.

The LH path with missing label $k=m+j$ starts by dropping
this label in~$Q$.
By (b), the label that is picked up in $Q$ is $\ell(j)$,
which is now duplicate, and the path proceeds by dropping
this label in $P$ which is the same as starting the Lemke
path on $\Pell$ with this missing label.
The LH path terminates by picking up the missing
label $k=m+j$ in $P$ by reaching the facet defined by
$(B\T x)_j=1$ which has label $\ell(j)$, so that the Lemke
path on $\Pell$ terminates.
Alternatively, label $k=m+j$ is picked up in $Q$ which by
(a) was preceded by dropping label $i=\ell(j)$, which was
duplicate because it was picked up in $P$ when encountering
the facet $x_i=0$, where $i$ is the missing label $\ell(j)$
on the Lemke path that has therefore terminated on $\Pell$.  
\endproof

\section{Hard-to-solve bimatrix games}
\label{s-hard}

With the help of Theorem~\ref{t-proj}, it suffices to
construct suitable labeled polytopes with (exponentially) long Lemke paths
in order to show that certain games have long LH paths.
McLennan and Tourky (2010) (summarized in Savani and von
Stengel 2006, Section~5) showed with the help of imitation
games that the polytopes with long Lemke paths due to Morris
(1994) can be used for this purpose.
In this section we extend this construction, with the help
of unit vector games, to games that are not square and that
are hard to solve not only with the Lemke--Howson algorithm,
but also with support enumeration methods.

In Section~\ref{s-perm} we present a very simple model
of random games that have very few Nash equilibria on
average, unlike games where all payoffs are chosen at
random.
These games are unit vector games, and the result
(Proposition~\ref{p-perm}) is joint work with Andy McLennan.
We then describe in Section~\ref{s-cyclic} dual cyclic
polytopes, whose facets have a nice combinatorial structure,
which have proved useful for the construction of games with
many equilibria, and with long LH paths.  
Our main result, Theorem~\ref{t-main} in Section~\ref{s-morris},
describes unit vector games based on dual cyclic polytopes
whose equilibria are hard to find not only with the LH
algorithm, but also with support enumeration.

\subsection{Permutation games}
\label{s-perm}

We present here a small ``warmup'' result that was found
jointly with Andy McLennan.
A \emph{permutation game} is an $n\times n$ game $(A,B)$
where $A$ is the identity matrix and $B$ is a permuted
identity matrix, that is, the $i$th row of $B$ is the unit
vector $e\T_{\pi(i)}$ for some permutation $\pi$ of
$\{1,\ldots,n\}$ (so column $\pi(i)$ is the best response to
row~$i$, and, because $A=I$, the best response to column $j$
is row~$j$).
Let $I^\pi$ be this matrix $B$, so that the permutation game
is $(I, I^\pi)$.

Because a permutation game $(I, I^\pi)$ is an imitation
game, the two strategies in an equilibrium have equal
support.
It is easy to see that any equilibrium of $(I, I^\pi)$
is of the form $(x,x)$ where $x$ mixes uniformly over its
support $S$ where $S$ is any nonempty subset of
$\{1,\ldots,n\}$ that is closed under $\pi$, that is,
$i\in S$ implies $\pi(i)\in S$.
In other words, $S$ is any nonempty union of \emph{cycles}
of~$\pi$.

A very simple model of a ``random'' game is to consider a
permutation game $(I,I^\pi)$ for a random permutation~$\pi$.

\begin{proposition}
\label{p-perm}
A random $n\times n$ permutation game has in expectation $n$
Nash equilibria.
\end{proposition}

\proof
Consider a random permutation $\pi$ of $\{1,\ldots,n\}$.
Let $E(n)$ be the expected number of Nash equilibria of 
$(I,I^\pi)$, where we want to prove that $E(n)=n$, which
is true for $n=1$.
Let $n>1$ and assume as inductive hypothesis that the claim
is true for $n-1$.
With probability $\frac1n$ we have $\pi(n)=n$,
in which case $\pi$ defines also
a random permutation of $\{1,\ldots,n-1\}$,
and any equilibrium of $(I,I^\pi)$ is
either the pure strategy equilibrium where both players play
$n$, or an equilibrium with a support $S$ of a random
$(n-1)\times (n-1)$ permutation game, or an equilibrium with
support $S\cup\{n\}$.
Hence, in this case the number of equilibria of $(I,I^\pi)$
is twice the number $E(n-1)$ of equilibria of a random
$(n-1)\times(n-1)$ game plus one.
Otherwise, with probability $\frac{n-1}n$, we have
$\pi(n)\ne n$, so that $\pi$ defines a 
random permutation of $\{1,\ldots,n-1\}$ when removing $n$
from the cycle of $\pi$ that contains~$n$.
For any equilibrium of the $(n-1)\times (n-1)$ permutation
game whose support contains this cycle, we add $n$ back to
the cycle to obtain the respective equilibrium of the
$n\times n$ game.
So in the case $\pi(n)\ne n$ the expected number of
equilibria is $E(n-1)$.
That is, 
\[
E(n)=\frac1n(1+2\cdot E(n-1))+\frac{n-1}n E(n-1) =
\frac1n(1+2(n-1)+(n-1)(n-1))%=\frac{n^2}n
=n,
\]
which completes the induction.  
\endproof

Random permutation games have very \emph{few} equilibria, as
Proposition~\ref{p-perm} shows.
In contrast, McLennan and Berg (2005) have shown that the
expected number of equilibria of an $n\times n$ game with
random payoffs is exponential in~$n$.
B\'ar\'any, Vempala, and Vetta (2007) show that such a game
has with high probability an equilibrium with small support.
A permutation game $(I,I^\pi)$, where the permutation $\pi$
has $k$ cycles, has $2^k-1$ many equilibria, but a large
number $k$ of cycles is rare.
In fact, there are $(n-1)!$ single-cycle permutations, so
%(one for each cycle of length $n$ which determines the
%permutation once its first element, say~$1$, is fixed), so
with probability $1/n=(n-1)!/n!$ the permutation game has
only a single equilibrium with full support.
For such games, an algorithm that enumerates all possible
supports starting with those of small size takes exponential
time.
% Hence, permutation games could be standard ``benchmark''
% games for equilibrium finding algorithms.
On the other hand, it is easy to see that the LH algorithm
finds an equilibrium in the shortest possible time, because
it just adds the strategies in a cycle of $\pi$ to its
current support.

However, a square game has only one full support, which is
natural to test as to whether it defines a (completely
mixed) equilibrium.
The full support always defines an equilibrium in a
permutation game.
It also does for the square games described by Savani and
von Stengel (2006) which have exponentially long LH paths.
They therefore constructed also non-square games where
support enumeration takes exponentially long time on
average.
It is an open question whether non-square games can be
constructed from unit vectors as an extension of permutation
games that are also hard to solve with support enumeration.  

\subsection{Cyclic polytopes and Gale evenness bitstrings}
\label{s-cyclic}

With the polytopes $P$ and $Q$ in (\ref{PQ}), Nash
equilibria of bimatrix games correspond to completely
labeled points of $P\times Q$.
The ``dual cyclic polytopes''
have the property that they have the maximal possible number
of vertices for a given dimension and number of facets
(see Ziegler 1995, or Gr\"un\-baum 2003).
In addition, it is easy to describe each vertex by the
facets it lies on.
Using these polytopes, von Stengel (1999) constructed
counterexamples for $n\ge 6$ to a conjecture by Quint and
Shubik (1997) that a nondegenerate $n\times n$ game has at
most $2^n-1$ equilibria.
McLennan and Park (1999) proved this conjecture for $n=4$;
the case $n=5$ is still open.
Morris (1994) gave a construction of labeled dual cyclic
polytopes with exponentially long Lemke paths, which we
extend in Theorem~\ref{t-main} below.

A standard way to define a \emph{cyclic polytope} $\Pprime$
in dimension~$m$ with $f$ vertices is as the convex hull of
$f$ points $\mu(t_j)$ on the \emph{moment curve} $\mu\colon
t\mapsto (t,t^2,\ldots,t^m)^\top$ for $1\le j\le f$.
However, the polytopes in (\ref{PQ}) are defined by
inequalities and not as convex hulls of points.
In the \emph{dual\/} (or ``polar'') of a polytope, its
vertices are reinterpreted as normal vectors of facets.
The polytope $\Pprime$ is first translated so that it has the
origin $\0$ in its interior, for example by subtracting
the arithmetic mean $\overline \mu$ of the points
$\mu(t_j)$ from each such point.
The resulting vectors $\mu(t_j)-\overline \mu$
then define the \emph{dual cyclic polytope} in dimension $m$
with $f$ facets
\begin{equation}
\label{polarcyc}
\dualcyc=\{\,x\in\reals^m  \mid 
(\mu(t_j)-\overline \mu)^{\top}x\le 1, ~1\le j\le f\,\}.
\end{equation} 
A suitable affine transformation of $\dualcyc$ (see von
Stengel 1999, p.~560) gives a polytope $P$ as in (\ref{PQ})
or (\ref{Pell}) so that the first $m$ inequalities of $P$
have the form $x\ge\0$.
The last $n=f-m$ inequalities $B^\top x\le\1$ of $P$
then determine the $m\times n$ payoff matrix~$B$.
If the first $m$ inequalities have labels $1,\ldots,m$ and
the last $n$ inequalities have labels
$\ell(1),\ldots,\ell(n)$, then this defines a labeled
polytope $\Pell$ as in (\ref{Pell}) and a unit vector game
as in Proposition~\ref{p-unitv}.

A vertex $u$ of $\dualcyc$
% a dual cyclic polytope in dimension $m$ with $f$ facets
is characterized by the \emph{bitstring}
$u_1u_2\cdots u_{f}$ of length $f$, where the $j$th bit
$u_j$ indicates whether $u$ is on the $j$th facet ($u_j=1$)
or not ($u_j=0$).
The polytope is simple, so exactly $m$ bits
are~$1$, and the other $f-m$ bits are~$0$.
Assume (which is all that is needed) that $t_1< t_2< \cdots
< t_{f}$ when defining the $j$th facet of $\dualcyc$ by the
binding inequality $(\mu(t_j)-\overline \mu)\T x=1$ in
(\ref{polarcyc}).
As shown by Gale (1963), the vertices of $\dualcyc$ are
characterized by the bitstrings that fulfill the \emph{Gale
evenness} condition: 
A bitstring with exactly $m$ $1$s represents a vertex if
and only if in any substring of the form $01^s0$ the number
$s$ of $1$s is even, so it has no odd-length substrings of
the form $010$, $01110$, and so on
(the reason is that the two zeros $u_i=u_j=0$ at the end of
such an odd-length substring would represent two points
$\mu(t_i)$ and $\mu(t_j)$ on the moment curve that are on
opposite sides of the hyperplane through the points
$\mu(t_k)$ for $u_k=1$, so that this hyperplane cannot
define a facet of the cyclic polytope that is the convex
hull of all the points, and therefore does not correspond to
a vertex of the dual cyclic polytope).  
Initial substrings $1^s0$ and terminal substrings $01^t$ are
allowed to have an odd number $s$ or $t$ of $1$s.
We only consider \emph{even} dimensions~$m$, where $s$ and
$t$ can only be both odd and by a cyclic shift (``wrapping
around'') of the bitstring define an even-length substring
$01^t1^s0$, which shows the cyclic symmetry of the Gale
evenness condition.

Consider, for even $m$, the bitstrings of length $f$ with
$m$ $1$s that fulfill Gale evenness, and as before let
$n=f-m$.
One such string is $1^m0^n$, that is, 
$m$ $1$s followed by $n$ $0$s.
For the corresponding vertex of $\dualcyc$, the first $m$
inequalities are tight, and if we label them with
$1,\ldots,m$, then this defines the completely labeled
vertex that is mapped to~$\0$ in the affine map from
$\dualcyc$ to the polytope~$P$, which will be a labeled
polytope $\Pell$.
The last $n$ facets of $\Pell$ correspond to the last $n$
positions of the bitstring, and they have labels
$\ell(1),\ldots,\ell(n)$.
If we view $\ell$ as a string $\ell(1)\cdots\ell(n)$ of $n$
labels, each of which is an element of $\{1,\ldots,m\}$,
then these labels specify a labeled polytope.
A \emph{completely labeled} vertex corresponds to a
Gale evenness bitstring $u_1\ldots u_f$ with $f=m+n$ where the
positions $i$ so that $u_i=1$ have all $m$ labels,
the label being $i$ if $1\le i\le m$, and $\ell(j)$ if
$i=m+j$ for $1\le j\le n$.  
We call the resulting polytope $\C{\ell}$, so this is the
dual cyclic polytope $\dualcyc$ where $f=m+n$ and $n$ is the
length of the string~$\ell$ of the last $n$ facet labels,
mapped affinely to $\Pell$ as in (\ref{Pell}), with facet
labels as described.

\subsection{Triple Morris games}
\label{s-morris}

In the notation just introduced, Morris (1994) studied Lemke
paths on the labeled dual cyclic polytope $\C{\sigma}$,
which we call the \emph{Morris polytope}, for a string
$\sigma$ of $m$ labels defined as follows.
Let $\tau$ be the string $\tau(1)\cdots\tau(m)$ of
$m$ labels, which is $1324$ for $m=4$, $132546$ for $m=6$,
$13254768$ for $m=8$, and in general defined by
\begin{equation}
\label{tau}
\tau(1)=1,
\qquad \tau(i)=i+(-1)^i
\quad(2\le i\le m-1),
\qquad
\tau(m)=m,
\end{equation}
and let $\sigma$ be the string $\tau$ in reverse order,
that is,
\begin{equation}
\label{sigma}
\sigma(i)= \tau(m-i+1)
\quad(1\le i\le m),
\end{equation}
so $\sigma=4231$ for $m=4$, $\sigma=645231$ for $m=6$,
$\sigma=86745231$ for $m=8$, and so on.
%Inspired by this construction,
We define the \emph{triple
Morris polytope} as $\C{\sigma\tau\sigma}$, where the
concatenated string $\sigma\tau\sigma$ is a string of $3m$
labels, for example $645231132546645231$ if $m=6$.

\begin{figure}[hbt]
% ps file says bounding box 261 high; take * .18 ex
\includegraphics[height=46.98ex]{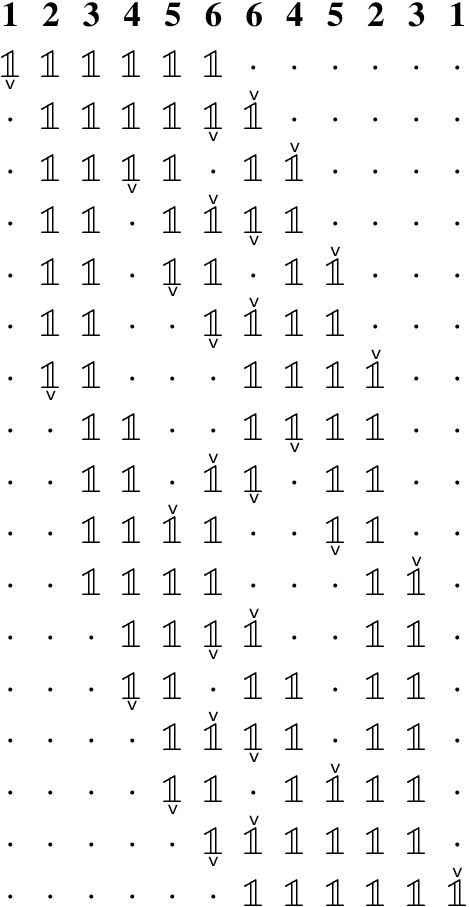}%
\hfill
%\vrule height 52mm ~ \vrule height 27ex
\hfill
\includegraphics[height=46.98ex]{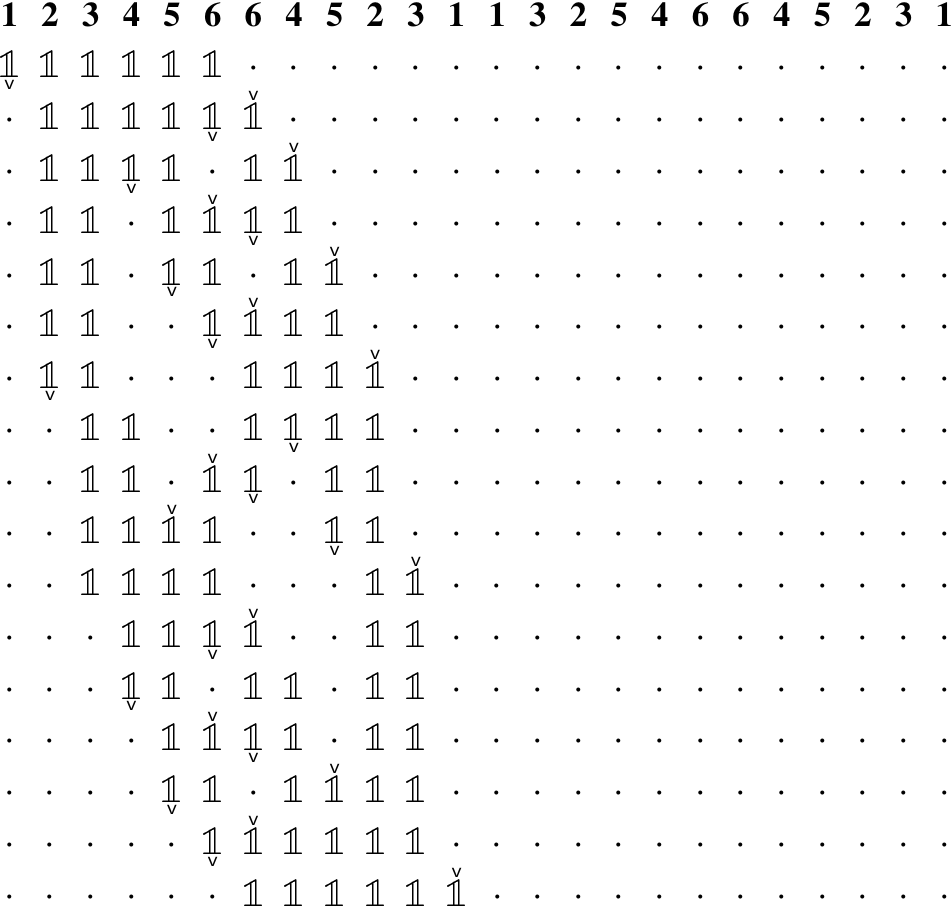}%
\caption{%
Lemke paths for missing label~$1$ on the Morris polytope
$\CM{\sigma}6$ (left), and on the triple Morris polytope
$\CM{\sigma\tau\sigma}6$ (right).%, for $m=6$.
}
\label{fpath1}
\end{figure}

The left-hand diagram of Figure~\ref{fpath1} shows the Lemke path for
missing label~$1$ on the Morris polytope $\C{\sigma}$ for
$m=6$.
The top row gives the labels, where the first $m$ are the
labels $1,\ldots,m$ corresponding to the inequalities
$x\ge\0$ in $\Pell$, followed by the labels $645231$ of
$\sigma$.
The rows below show the vertices of $\CM{\sigma}{6}$ as bitstrings,
where bit $1$ is written in a different font and
$0$ as a dot $\cdot$ to distinguish them better.
The first string $111111000000$ represents the starting
vertex $\0$ of $\Pell$.
A facet that is left by dropping a label, at first the
missing label~$1$, has a small ``\textsf{v}'' underneath the bit
$1$, whereas the facet that is just encountered, with the
corresponding label that is picked up, has the
``\textsf{v}'' above it.
Dropping label $1$ means the second vertex is
$011111100000$, where label $6$ is picked up and duplicate.
Because the previous facet with that label corresponds to
the second to last bit $1$, it is dropped next, which gives the
next vertex as $011110110000$ where label $4$ is picked up, and
so on.

The right-hand diagram in Figure~\ref{fpath1} shows the Lemke
path for missing label~$1$ on the triple Morris polytope 
$\CM{\sigma\tau\sigma}6$.
Because in this case the only affected bits are those with
labels in the first substring $\sigma$ of the entire label
string $\sigma\tau\sigma$, the path is essentially the same
as in the Morris polytope $\CM{\sigma}6$ on the left.

\begin{figure}[hbt]
% ps file says bounding box 142 high; take * .18 ex
\includegraphics[height=25.56ex]{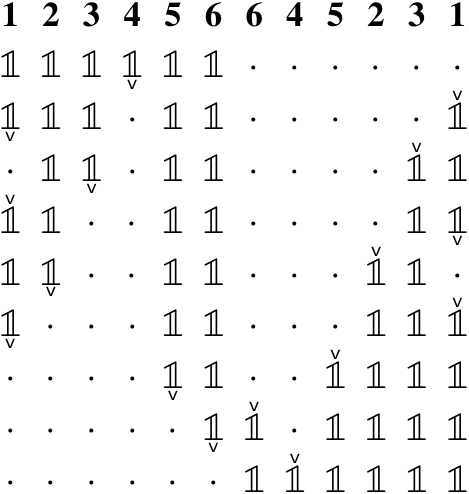}%
\hfill
%\vrule height 52mm ~ \vrule height 27ex
\hfill
\includegraphics[height=25.56ex]{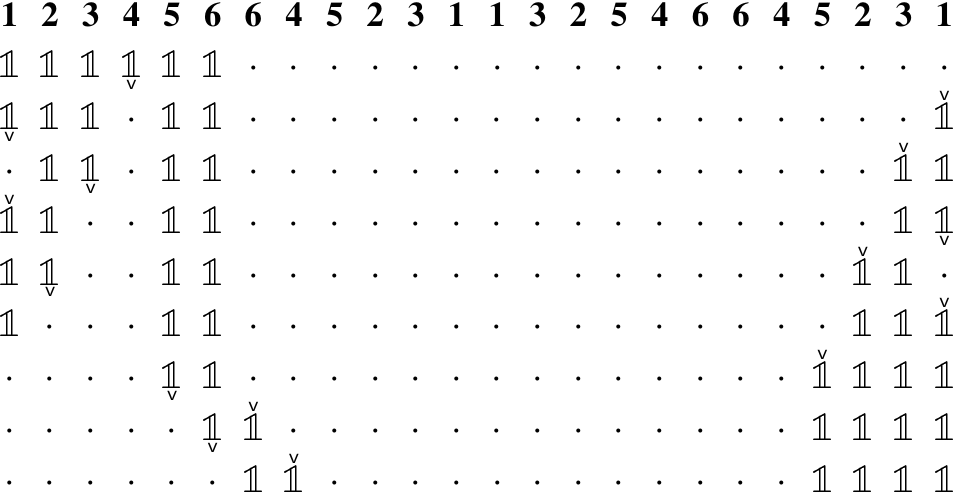}%
\caption{%
Lemke paths for missing label~$4$ on the Morris polytope
$\CM{\sigma}6$ (left), and on the triple Morris polytope
$\CM{\sigma\tau\sigma}6$ (right).%, for $m=6$.
}
\label{fpath4}
\end{figure}

Figure~\ref{fpath4} shows the Lemke paths for these two
polytopes for the missing label~$4$.
In this case, to preserve Gale evenness, the bitstring that
follows the starting bitstring $111111000000$ is
$111011000001$, which ``wraps around'' the left end to add a
bit $1$ in the rightmost position, which has label~$1$ that
is picked up.
The resulting path is the composition of two sub-paths.
The first path moves away from the dropped label $4$ to the
left (and wrapping around), which behaves essentially like a
path with dropped label $4$ on a Morris polytope in
dimension~$4$, until label~$5$ is picked up.
This starts the second sub-path with the original label~$5$
being dropped, which is essentially a (rather short) path
with dropped label~$1$ on a Morris polytope in
dimension~$2$.

In general, the Lemke path on the Morris polytope
$\CM{\sigma}{m}$ for the missing label $k$, where $k$ is
even, is the composition of two sub-paths.
The first sub-path is equivalent to a Lemke path for missing
label $k$ on a Morris polytope $\CM{\sigma}{k}$ for missing
label $k$, which, by symmetry (writing the label strings
backwards), is the same as the Lemke path on $\CM{\sigma}{k}$
for missing label $1$.
The second sub-path is equivalent to a Lemke path for
missing label $1$ on a Morris polytope $\CM{\sigma}{m-k}$.
(If $k$ is odd, then a similar consideration applies by
symmetry.)
In this way, the Lemke paths for any missing label $k$ are
described by considering Lemke paths for missing label~$1$
in dimension $k$ or $m-k$, where clearly $k$ or $m-k$ is at
least $m/2$ (which is used in the second part of
Theorem~\ref{t-morris} below).

The right-hand diagram in Figure~\ref{fpath4} shows that the same
Lemke path for missing label~$4$ results in the triple
Morris polytope $\CM{\sigma\tau\sigma}6$, because the two
copies of the label string $\sigma$ have the same effect as
the single label string $\sigma$ in $\CM{\sigma}6$.
Clearly, this correspondence holds for any missing label.

\begin{proposition}
\label{p-corpaths}
There is a one-to-one correspondence between 
the Lemke path for missing label $k$ starting from the
Gale evenness string$1^m0^m$ (vertex $\0$) of the Morris
polytope $\C{\sigma}$
and the Lemke path for missing label $k$ starting from the
Gale evenness string $1^m0^{3m}$ (vertex $\0$) of
the triple Morris polytope $\C{\sigma\tau\sigma}$,
for $1\le k\le m$.  
\end{proposition}

The length of the Lemke path for missing label $1$ on
$\C{\sigma}$ is \emph{exponential} in the dimension~$m$.
Essentially, this path composed of two such paths in
dimension~$m-2$, with another such path in dimension~$m-4$
between them (Figure~\ref{fpath1} gives an indication).
Hence, if the length of the path is $a_m$, the recurrence
$a_m=2a_{m-2}+a_{m-4}$ implies that it grows from $a_{m-2}$
to $a_m$ by an approximate factor of $1+\sqrt2$;
for details see Morris (1994), and for similar arguments
Savani and von Stengel (2006, Theorem~7).
Recall that $\Theta(f(n))$ means bounded
above and below by a constant times $f(n)$ for large~$n$.

\begin{theorem}{}\emph{(Morris 1994, Proposition~3.4)}
\label{t-morris}
The longest Lemke path on $\C{\sigma}$ is for missing
label~$1$ and has length $\Theta((1+\sqrt2)^{m/2})$.
The shortest Lemke path on $\C{\sigma}$ is for missing 
label~$m/2$ and has length $\Theta((1+\sqrt2)^{m/4})$.  
\end{theorem}

Consequently, the Lemke paths on triple Morris polytopes are
also exponentially long.
Hence, these polytopes define unit vector games which by
Theorem~\ref{t-proj} have exponentially long LH paths.
We consider these games because they are of dimension
$m\times 3m$ rather than $m\times m$ for the unit vector
game defined by the Morris polytope $\C{\sigma}$.  
The latter, square game has a single completely mixed
equilibrium, which is easily found by support enumeration.
We show next that the $m\times 3m$ game has multiple
equilibria, each of them with full support for player~1 
(for which we need the ``middle'' label string $\tau$).

\begin{proposition}
\label{p-3eq}
The $m\times 3m$ unit vector game that corresponds to the
triple Morris polytope $\C{\sigma\tau\sigma}$ has $3^{m/2}$
Nash equilibria.
Each of them has full support for player~$1$.
\end{proposition}

\begin{figure}[hbt]
\strut
\hfill
\includegraphics[height=35.5ex]{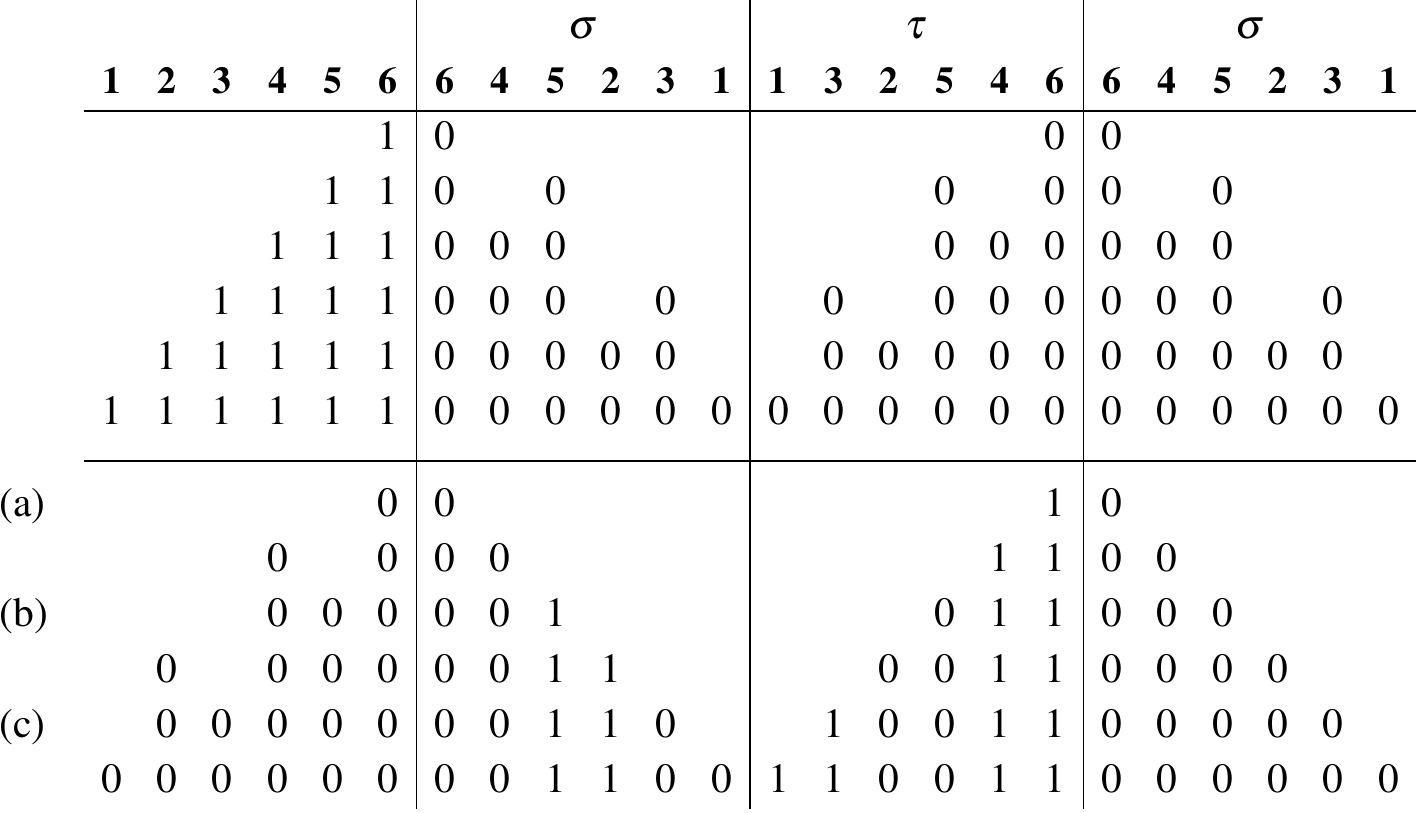}%
%\vrule height 52mm ~ \vrule height 27ex
\hfill
\strut
\caption{%
Illustration of the proof of Proposition \ref{p-3eq} for $m=6$.
The top half shows the only completely labeled
bitstring $u=1^m0^{3m}$ where $u_m=1$,
the bottom half one such string where $u_m=0$.
There are three choices in each of the $m/2$ lines (a), (b), (c).
}
\label{fexplain}
\end{figure}

\proof
With the label string $\sigma\tau\sigma$, we identify the
completely labeled vertices of $\C{\sigma\tau\sigma}$ as
completely labeled Gale evenness strings
$u=u_1\cdots u_{4m}$.  
First, we show that if $u_m=1$, then $u=1^m0^{3m}$, which is
the vertex $\0$. 
This is illustrated in the top part of Figure~\ref{fexplain}.
Because $u_m=1$ which has label $m$, the other positions $j$
with label $m$ have $u_j=0$, which are $j=m+1$ and $j=3m+1$
(the first positions in the two substrings $\sigma$ of 
$\sigma\tau\sigma$) and $j=3m$ (the last position of
$\tau$), shown in the first line of Figure~\ref{fexplain}.
The substring $10$ with bits $u_m$ and $u_{m-1}$ requires
$u_{m-1}=1$ by Gale evenness, with label $m-1$, which now
requires $u_j=0$ for $j=m+3,3m-2,3m+3$, as shown for
label~$5$ in the second line of Figure~\ref{fexplain}.
The single bit $u_{m+2}$ (label~$4$ in the picture) must be
$0$ by Gale evenness, and similarly $u_{3m-1}=u_{3m+2}=0$
and hence $u_{m-2}=1$, as shown in the next line.
Continuing in this manner, the only possible string is
$u=1^m0^n$ as claimed.

Suppose now that $u_m=0$, where we will show that the
resulting completely labeled Gale evenness string is of the
form $u=0^m\beta$, which represents a Nash equilibrium
$(x,y)$ of the game with full support for player~1, that is,
$x>\0$.
Consider the lower part of Figure~\ref{fexplain}, where in
the first line (a) we now have three choices where to put
the label $m$, namely by setting $u_j=1$ for exactly one $j$
in $\{m+1,3m,3m+1\}$, corresponding to the first position in
one of the $\sigma$s or the last position in $\tau$ (where
we choose the latter in the picture).
So $u_{3m}=1$, which requires $u_{3m-1}=1$ by Gale evenness
because $u_{3m+1}=0$.
This next position always has label $m-2$ (label $4$ if $m=6$),
so that $u_{m-2}=0$, and similarly in the other positions
with that label.
But then $u_{m-1}=0$ by Gale evenness and we again have 
three choices, in one of the substrings $\sigma$, $\tau$,
$\sigma$, of where to set that next bit with label $m-1$.
In the picture, we choose it in line (b) in the first
substring $\sigma$, that is, $u_{m+3}=1$.
Continuing in that manner, there are $m/2$ times where we
can choose a pair $11$ of two bits $1$ in either substring
$\sigma$, $\tau$, $\sigma$ to obtain a completely labeled
Gale evenness string, making $3^{m/2}$ choices in total,
as claimed.  
\endproof

The $m\times 3m$ game in Proposition~\ref{p-3eq} has an
exponential number of equilibria, which define a certain set
$E$ of equilibrium supports of player~2.
However, they form an exponentially small subset of all
possible supports.
An equilibrium is therefore hard to find with a support
enumeration algorithm, even if that algorithm is restricted
to testing only supports of size $m$ for player~2.

\begin{proposition}
\label{p-hide}
Consider an $m\times 3m$ game where a pair of supports
defines a Nash equilibrium if and only if both supports have
size $m$, and player~2's support belongs to the set $E$,
a set of $m$-sized subsets of $\{1,\ldots,3m\}$.
A support enumeration algorithm %, even if it knows~$E$,
that tests supports picked uniformly at random without
replacement from the set $U$
of all $m$-sized subsets of $\{1,\ldots,3m\}$
has to test an expected number of
\begin{equation}
\label{numguess}
\frac{\binom{3m}{m}-|E|}{|E|+1}+1
\end{equation}
supports before finding an equilibrium support.
\end{proposition}

\proof
To find the expected number of guesses required to find an
equilibrium we use a standard argument 
(Motwani and Raghavan 1995, p.~10). 
Consider a random enumeration of the elements of~$U$. 
The elements of $U-E$, which we index by $i= 1,\ldots,
{|U-E|}$, correspond to non-equilibrium supports.
Let $W_i$ be the \emph{indicator variable} that 
takes value $1$ if the $i$th element of
${U-E}$ precedes all members of $E$ in the enumeration 
of $U$, and $0$ otherwise.
Then $W=\sum_{i=1}^{|U-E|}{W_i}$ is the random variable 
equal to the number of supports checked before the first 
equilibrium is found.
For a single element of $U-E$, the probability that it is
in front of all elements of $E$ is $1/(|E|+1)$.
Hence, using the linearity of expectation, 
\[
\E(W)=\E(\sum_{i=1}^{|U-E|}{W_i})
=\sum_{i=1}^{|U-E|}\E({W_i})
=\sum_{i=1}^{{|U-E|}}\frac{1}{|E|+1}
=\frac{{|U|-|E|}}{|E|+1}
=\frac{{\binom{3m}{m}-|E|}}{|E|+1}.
\] 
This shows that the expected number of support guesses until
an equilibrium is found is given by \eqref{numguess}, as
claimed. 
\endproof

In Proposition~\ref{p-hide}, we assume that the algorithm
does not identify any particular pattern as to which
supports should be tested.
One way to achieve this is to permute the columns of the
game randomly
(if one knows that the payoff matrix $B$ of player~2 is derived
from a dual cyclic polytope, then this random order can be
identified with a specialized method, see Savani 2006,
Section 3.6; this is not a general method for solving games
so we do not consider it).
However, unless one distorts the polytope $Q$ in
(\ref{QNi}), this still leaves a payoff matrix of player~1
where each unit vector appears three times.
In this case, even if the algorithm picks only columns where each
unit vector appears once, there would be $3^m$ possible
supports which define a set $U$ of size $3^m$ rather than
$\binom{3m}{m}$ in Proposition~\ref{p-hide}.
Such a set $U$ is still exponentially large compared to the
set $E$ of $3^{m/2}$ supports that define a Nash
equilibrium.
In that case the expected time for the support-testing
algorithm in the following theorem is $(\sqrt3)^m\approx
1.732^m$.

\begin{theorem}
\label{t-main}
Finding a Nash equilibrium of the $m\times 3m$ unit-vector
game that corresponds to the triple Morris polytope
$\C{\sigma\tau\sigma}$ takes at least time
$\Theta((1+\sqrt2)^{m/4}) \approx\Theta(1.246^m)$ with the
Lemke--Howson algorithm, 
and on expectation time $\Theta((27/4\sqrt3)^m/\sqrt m)
\approx\Theta(3.897^m/\sqrt m)$ with
an algorithm that tests in random order arbitrary supports of size
$m\times m$ of the game.  
\end{theorem}

\proof
The length of the LH paths follows from
Theorem~\ref{t-morris}, Proposition~\ref{p-corpaths}, and
Theorem~\ref{t-proj}.
For the support-testing algorithm, we have $|E|= (\sqrt3)^m$
in Proposition~\ref{p-hide} by Proposition~\ref{p-3eq}.
Using Stirling's formula $n!\sim \sqrt{2\pi n}\cdot(n/e)^n$,
we have $\binom{3m}{m}\sim(\sqrt 3\cdot 3^{3m})/(2 \sqrt{\pi
m}\cdot 2^{2m})$, so that the expression in (\ref{numguess})
is $\Theta((27/4\sqrt3)^m/\sqrt m)$.
\endproof

To conclude, we note results on the following combinatorial
problem: 
Let $m$ be even and let $\ell$ be a string of $n$ labels
from $\{1,\ldots,m\}$, and consider the set of Gale evenness
bitstrings of length $m+n$ which encode the vertices of the
labeled polytope $\C{\ell}$.
The problem is to find a second completely labeled Gale
evenness string other than $1^m0^n$.
Casetti, Merschen, and von Stengel (2010) have shown that
this is equivalent to finding a second perfect matching in
the \emph{Euler graph} with nodes $1,\ldots,m$ and edges defined by
the Euler tour $1,\ldots,m,\ell(1),\ldots,\ell(n),1$.
The edges in a perfect matching encode the pairs of $1$s in
a Gale evenness bitstring, which is completely labeled
because the edges cover all nodes.
V\'egh and von Stengel (2015, Theorem~12) give a near-linear
time algorithm that finds such a second perfect matching
that, in addition, has opposite \emph{sign}, which
corresponds to a Nash equilibrium of positive index as it
would be found by a Lemke path (which, however, can be
exponentially long).  
So this combinatorial problem is simpler than the problem of
finding a Nash equilibrium of a bimatrix game, even though
it gives rise to games that are hard to solve by the
standard methods considered in Theorem~\ref{t-main}.

\section*{Acknowledgements}

The first author was supported by EPSRC grant EP/L011018/1.
We thank the referees for helpful comments.  

\section*{References}
\addcontentsline{toc}{section}{References} 
\frenchspacing
\parindent=-2em\advance\leftskip by2em
\parskip=.3ex minus .1ex
\footnotesize
\small
\hskip\parindent
Balthasar, A. V. (2009),
\textit{Geometry and Equilibria in Bimatrix Games},
PhD Thesis, London School of Economics.

B\'ar\'any, I., S. Vempala, and A. Vetta (2007),
``Nash equilibria in random games,''
\textit{Random Structures and Algorithms} 31, 391--405.

Casetti, M. M., J. Merschen, and B. von Stengel (2010),
``Finding Gale strings,''
\textit{Electronic Notes in Discrete Mathematics} 36, 1065--1072.

Chen, X., and X. Deng (2006),
``Settling the complexity of two-player Nash equilibrium,''
\textit{Proc. 47th Symp. Foundations of Computer Science
(FOCS)}, 261--272.  

% Codenotti, B., S. De Rossi, and M. Pagan (2008),
% An experimental analysis of Lemke-Howson algorithm.
% arXiv preprint 0811.3247.

% Cottle, R. W., and G. B. Dantzig (1968),
% Complementary pivot theory of mathematical programming.
% Linear Algebra and Its Applications 1, 103--125.

Cottle, R. W., and G. B. Dantzig (1970),
``A generalization of the linear complementarity problem,''
\textit{J. Combinatorial Theory} 8, 79--90.

Cottle, R. W., J.-S. Pang, and R. E. Stone (1992),
\textit{The Linear Complementarity Problem}. Academic Press,
San Diego.

Daskalakis, C., P. W. Goldberg, and C. H. Papadimitriou
(2009),
``The complexity of computing a Nash equilibrium,''
SIAM Journal on Computing 39, 195--259.  

Dickhaut, J., and T. Kaplan (1991),
``A program for finding Nash equilibria,''
\textit{The Mathematica Journal} 1, Issue 4, 87--93.  
%%% Note to copy editor: Keep "Issue" as this journal
%%% numbers pages separately per issue

Gale, D. (1963),
``Neighborly and cyclic polytopes,'' V.  Klee, ed.,
\textit{Convexity, Proc. Seventh Symposium in
Pure Mathematics, Vol. 7}, 225--232,
Providence, Rhode Island: American Mathematical Society.  

Gale, D., H. W. Kuhn, and A. W. Tucker (1950),
``On symmetric games,''
H.~W. Kuhn and A.~W. Tucker, eds.,
\textit{Contributions to the Theory of Games I,
Annals of Mathematics Studies 24}, 81--87,
Princeton: Princeton University Press.

Goldberg, P. W., C. H. Papadimitriou, and R. Savani (2013),
``The complexity of the homotopy method, equilibrium
selection, and Lemke--Howson solutions,''
\textit{ACM Transactions on Economics and Computation} 1, Article 9.

Gr\"unbaum, B. (2003),
\textit{Convex Polytopes, 2nd ed.},
Graduate Texts in Mathematics, Vol. 221,
New York: Springer.

Lemke, C. E. (1965),
``Bimatrix equilibrium points and mathematical
programming,''
\textit{Management Science} 11, 681--689.

Lemke, C. E., and J. T. Howson, Jr. (1964),
``Equilibrium points of bimatrix games,''
\textit{Journal of the Society for Industrial and
Applied Mathematics} 12, 413--423.

McLennan, A., and J. Berg (2005),
``Asymptotic expected number of Nash equilibria of two-player
normal form games,''
\textit{Games and Economic Behavior} 51, 264--295.

McLennan, A., and I.-U. Park (1999),
``Generic $4\times 4$ two person games have at most 15 Nash
equilibria,''
\textit{Games and Economic Behavior} 26, 111--130.

% McLennan, A., and R. Tourky (2010),
% Simple complexity from imitation games.
% Games and Economic Behavior 68, 683--688.

McLennan, A., and R. Tourky (2010),
``Imitation games and computation,''
\textit{Games and Economic Behavior} 70, 4--11.  

Morris, W. D., Jr. (1994),
``Lemke paths on simple polytopes,''
\textit{Mathematics of Operations Research} 19, 780--789.

Motwani, R., and P. Raghavan (1995),
\textit{Randomized Algorithms},
Cambridge: Cambridge University Press.

% Nash, J. (1951),
% Non-cooperative games.
% Annals of Mathematics 54, 286--295.  

Papadimitriou, C. H. (1994),
``On the complexity of the parity argument and other
inefficient proofs of existence,''
\textit{Journal of Computer and System Sciences} 48, 498--532.  

Quint, T., and M. Shubik (1997),
``A theorem on the number of Nash equilibria in a bimatrix
game,''
\textit{International Journal of Game Theory} 26, 353--359.  

Savani, R. (2006),
\textit{Finding Nash Equilibria of Bimatrix Games},
PhD Thesis, London School of Economics. 

Savani, R., and B. von Stengel (2006),
``Hard-to-solve bimatrix games,''
\textit{Econometrica} 74, 397--429.  

Shapley, L. S. (1974),
``A note on the Lemke--Howson algorithm,''
\textit{Mathematical Programming Study 1: Pivoting and
Extensions}, 175--189.  

V\'egh, L. A., and B. von Stengel (2015),
``Oriented Euler complexes and signed perfect matchings,''
\textit{Mathematical Programming Series B} 150, 153--178. 

von~Stengel, B. (1999),
``New maximal numbers of equilibria in bimatrix games,''
\textit{Discrete and Computational Geometry} 21, 557--568.

von~Stengel, B. (2002),
``Computing equilibria for two-person games,''
R. J. Aumann and S. Hart, eds.,
\textit{Handbook of Game Theory, Vol.~3},
1723--1759, Amsterdam: North-Holland.

Ziegler, G. M. (1995),
\textit{Lectures on Polytopes}.
Graduate Texts in Mathematics, Vol. 152, 
New York: Springer.  

\end{document}